\documentclass[%
 reprint,
superscriptaddress,
 amsmath,amssymb,
 aps,
pra,
]{revtex4-2}

\DeclareMathOperator{\sinc}{sinc}
\DeclareMathOperator{\erfc}{erfc}

\usepackage{graphicx}
\usepackage{bm}
\usepackage{hyperref}
\usepackage{wrapfig}
\hypersetup{
    colorlinks=true,
    linkcolor=blue,
    filecolor=blue,      
    urlcolor=blue,
    citecolor=cyan,
    }

\begin{document}
\title{Time-resolved second-order autocorrelation function of parametric downconversion}

\author{Dmitri B. Horoshko}\email{dmitri.horoshko@uni-ulm.de}
\affiliation{Univ. Lille, CNRS, UMR 8523 - PhLAM - Physique des Lasers Atomes et Mol\'{e}cules, F-59000 Lille, France}
\affiliation{Institut f\"ur Quantenoptik, Universit\"at Ulm, Ulm D-89073, Germany}
\author{Shivang Srivastava}
\affiliation{Univ. Lille, CNRS, UMR 8523 - PhLAM - Physique des Lasers Atomes et Mol\'{e}cules, F-59000 Lille, France}
\affiliation{Telecom Paris, Institut Polytechnique de Paris, 19 Place Marguerite Perey, 91120 Palaiseau, France}
\author{Filip So{\'s}nicki}
\affiliation{Faculty of Physics, University of Warsaw, Pasteura 5, 02-093 Warszawa, Poland}
\affiliation{Paderborn University, Department of Physics, Integrated Quantum Optics, Institute for Photonic Quantum Systems (PhoQS), Warburger Straße 100, 33098 Paderborn, Germany} 
\author{Micha{\l} Miko{\l}ajczyk}
\affiliation{Faculty of Physics, University of Warsaw, Pasteura 5, 02-093 Warszawa, Poland}
\author{Micha{\l} Karpi\'nski}
\affiliation{Faculty of Physics, University of Warsaw, Pasteura 5, 02-093 Warszawa, Poland}
\author{Benjamin Brecht}
\affiliation{Paderborn University, Department of Physics, Integrated Quantum Optics, Institute for Photonic Quantum Systems (PhoQS), Warburger Straße 100, 33098 Paderborn, Germany}
\author{Mikhail I. Kolobov}
\affiliation{Univ. Lille, CNRS, UMR 8523 - PhLAM - Physique des Lasers Atomes et Mol\'{e}cules, F-59000 Lille, France}
\date{\today}

\begin{abstract}
We study a possibility of measuring the time-resolved second-order autocorrelation function of one of two beams generated in type-II parametric downconversion by means of temporal magnification of this beam, bringing its correlation time from the picosecond to the nanosecond scale, which can be resolved by modern photodetectors. We show that such a measurement enables one to infer directly the degree of global coherence of that beam, which is linked by a simple relation to the number of modes characterizing the entanglement between the two generated beams. We illustrate the proposed method by an example of photon pairs generated in a periodically poled KTP crystal with a symmetric group velocity matching for various durations of the pump pulse, resulting in different numbers of modes. Our theoretical model also shows that the magnified double-heralded autocorrelation function of one beam exhibits a local maximum around zero delay time, corresponding to photon bunching at a short time scale.
\end{abstract}
\maketitle

\section{Introduction}
The second-order intensity autocorrelation function of the optical field, $g^{(2)}(\tau)$, where $\tau$ is a time delay, is a powerful tool to determine the physical properties of light beams and their sources \cite{Glauber63,Mandel&Wolf}. In particular, the value of this function at $\tau=0$ reveals the statistics of the photons: $g^{(2)}(0)=2$ for Gaussian statistics, $g^{(2)}(0)=1$ for Poissonian one, and the value $g^{(2)}(0)<g^{(2)}(\tau)$, where $\tau>0$, indicates the phenomenon of photon antibunching, which is a signature of field nonclassicality \cite{Mandel&Wolf}. Measurement of this function with fully resolved temporal dependence became in the last decade a routine task in the characterization of single-photon sources such as quantum dots \cite{Arakawa20}, single molecules \cite{Toninelli21}, and diamond color centers \cite{Doherty13}, where its width is in the nanosecond scale, above the resolution time of modern photodetectors ranging from about 50 ps. Another important source of single photons and photon pairs, single-pass parametric downconversion (PDC), is, however, hard to characterize in this way, because of a rather high bandwidth, more than 1 THz, of the generated light, placing the typical width of the $g^{(2)}(\tau)$ function into the sub-picosecond range, well below the photodetector resolution time. A time-integrated second-order autocorrelation function $g^{(2)}_\text{int}$ was introduced for this case \cite{Christ11}, which in particular allows one to determine the number of modes $K$ of two entangled beams by measuring the autocorrelation function of one of them: $K=(g^{(2)}_\text{int}-1)^{-1}$. However, in a highly multimode case, where $g^{(2)}_\text{int}\to1$, the evaluated number of modes becomes highly sensitive to the measurement error in determining $g^{(2)}_\text{int}$. 

A time-resolved measurement of the second-order autocorrelation function for a broadband field is possible with the help of a temporal imaging technique, which allows one to stretch or compress optical waveforms in time \cite{Kolner94,Salem13}. This technique, known for several decades in classical optics, has been successfully applied to nonclassical fields \cite{Lavoie13,Karpinski17,Mittal17} and the corresponding quantum theory of this technique has been developed \cite{Patera17,Patera18,Shi20}. Recent experiments with single photons \cite{Sosnicki20,Joshi22,Sosnicki23} demonstrated the possibility of bringing their picosecond-scale temporal features to the nanosecond scale, thus paving the way for a time-resolved measurement of the $g^{(2)}(\tau)$ function. In this paper, we demonstrate how the temporal width of the second-order autocorrelation function measured in this way can be used to precisely determine the number of modes in PDC. We also analyze the heralded version of the $g^{(2)}(\tau)$ function and predict that it has a local maximum at short times around zero, which is a signature of photon bunching. 

The paper is structured in the following way. In Sec. \ref{sec:Framework}, we describe the general framework for generation of photon pairs in a nonlinear crystal, their temporal magnification, and detection. In Sec. \ref{sec:calculation}, we give an analytical expression for the second-order autocorrelation function in the low-gain regime of PDC with a Gaussian model of the phase-matching function. We illustrate the different limiting cases of the general formalism by considering a symmetric group velocity matching in a periodically poled KTP crystal. We extend the developed theory to the case of heralded photons in Sec.~\ref{sec:Heralded}. Section~\ref{sec:Conclusion} concludes the paper.

\begin{figure*}[ht]
\centering
\includegraphics[width=18cm]{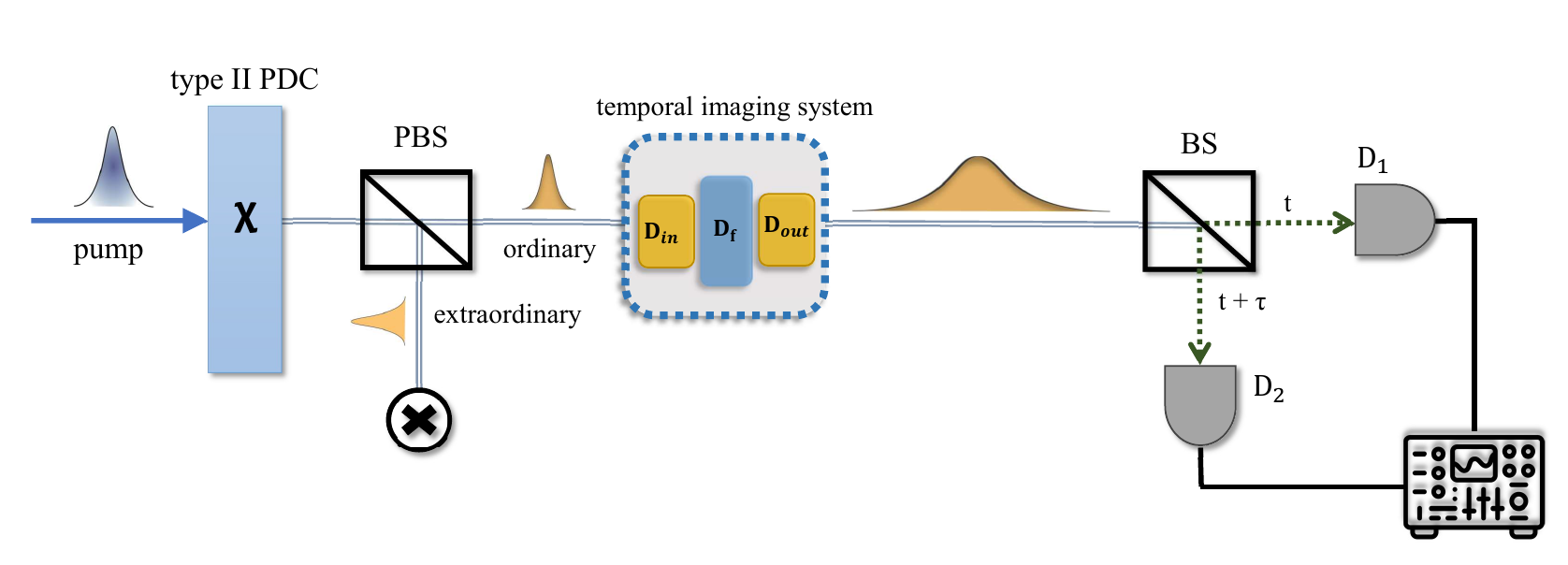}
\caption{Scheme for generation of photon pairs and detection of their second-order correlation function. A pump pulse impinges on a nonlinear crystal cut for type-II collinear PDC. Two subharmonic pulses appear as ordinary and extraordinary waves in the crystal. The pump is removed, while
two subharmonic pulses are separated by a polarizing beam splitter (PBS). The extraordinary pulse is discarded. The ordinary pulse passes through a temporal imaging system, composed of an input dispersive medium, a time lens, and an output dispersive medium. The stretched ordinary pulse is split by a beam splitter (BS) and detected by photodetectors $D_1$ and $D_2$. The second-order autocorrelation function of the ordinary wave is obtained from the records of both photodetectors.\label{fig:scheme}}
\end{figure*}

\section{General framework \label{sec:Framework}}

The scheme for the generation of two entangled beams of light in PDC and detection of the autocorrelation function of one of them is shown in Fig. \ref{fig:scheme}. Each of its components is explained and modeled in subsequent sections. In contrast to a more traditional approach to the description of low-gain PDC in the Schr\"odinger picture \cite{KlyshkoBook,Rubin94,Grice97,Keller97}, we describe the quantum optical field transformation in the Heisenberg picture to ensure a natural extension to the high-gain regime and because the quantum temporal imaging formalism is developed in the Heisenberg picture \cite{Patera18,Shi20}. 

\subsection{Parametric downconversion \label{sec:PDC}}

We consider a parametric downconversion in a second-order nonlinear crystal with a type-II collinear phase-matching. The plain-wave pump with central frequency $\omega_p$ propagates along the $z$ axis and is polarized as an ordinary or extraordinary wave. As a result of the nonlinear transformation of the pump field in the crystal, two subharmonic waves with the central frequencies $\omega_o$ and $\omega_e$, summing up to the pump frequency, $\omega_o+\omega_e=\omega_p$, emerge, polarized as ordinary and extraordinary waves. The interaction of these waves is most easily described in the Heisenberg picture in the spectral domain. The positive-frequency part of the field of each wave is \cite{Kolobov99}
\begin{equation}\label{FourierSignal}
\hat E_\mu^{(+)}(z,t) = i\mathcal{E}_\mu\int  \epsilon_\mu(z,\Omega)e^{ik_\mu(\Omega)z-i(\omega_\mu+\Omega)t} \frac{d\Omega}{2\pi},		
\end{equation}
where $\mu$ takes values $\{p,o,e\}$ for the pump, ordinary and extraordinary subharmonic waves respectively, $t$ is time, $\Omega$ denotes the frequency detuning from the carrier frequency, $k_\mu(\Omega)$ is the wave vector of the corresponding wave at frequency $\omega_\mu+\Omega$, and $\mathcal{E}_\mu$ is the single-photon electric field
\begin{equation}
\mathcal{E}_\mu = \left(\frac{\hbar\omega_\mu}{2\varepsilon_0c\mathcal{A}n_\mu}\right)^{\frac12}
\end{equation}
with $\varepsilon_0$ the vacuum permittivity, $\mathcal{A}$ the cross-sectional area of the light beam, and $n_\mu$ the refractive index of the corresponding wave. For a strong undepleted pump, its spectral amplitude $\epsilon_p(\Omega,z)$ is a $c$-number independent of $z$. We assume that it is a Gaussian transform-limited pulse $E_p^{(+)}(0,t) = E_\text{peak}e^{-t^2/4\Delta t^2-i\omega_pt}$ with a full width at half maximum (FWHM) duration $\tau_p=2\sqrt{2\ln2}\Delta t$. From Eq. (\ref{FourierSignal}), we obtain $\epsilon_p(0,\Omega) = \alpha(\Omega)=\alpha_0\exp(-\Omega^2/4\Omega_p^2)$, where $\Omega_p=\sqrt{2\ln 2}/\tau_p$ and $\alpha_0=-i2\sqrt{\pi}\Delta t E_\text{peak}/\mathcal{E}_p$.

The spectral amplitude of the ordinary or extraordinary subharmonic wave, $\epsilon_\mu(z,\Omega)$, is the annihilation operator of a photon at position $z$ with the frequency $\omega_\mu+\Omega$ and the corresponding polarization, satisfying the canonical equal-space commutation relations \cite{Huttner90,Kolobov99,Horoshko22} $\left[\epsilon_\mu(z,\Omega),\epsilon_\nu^\dagger(z,\Omega')\right]
= 2\pi\delta_{\mu\nu}\delta(\Omega-\Omega')$. The evolution of this operator along the crystal is described by the spatial Heisenberg equation \cite{Shen67}
\begin{equation}\label{evolution}
    \frac{d}{dz}\epsilon_\mu(z,\Omega) = \frac{i}\hbar\left[\epsilon_\mu(z,\Omega),G(z)\right],
\end{equation}
where the spatial Hamiltonian $G(z)$ is given by the momentum transferred through the plane $z$ \cite{Horoshko22} and equals
\begin{equation}\label{G}
    G(z) = \chi(z)\int\limits_{-\infty}^{+\infty} E^{(+)}_p(z,t)\hat E^{(-)}_o(z,t)\hat E^{(-)}_e(z,t)dt + \mathrm{H.c.},
\end{equation}
where $\hat E^{(-)}_\mu(z,t)=\hat E^{(+)\dagger}_\mu(z,t)$ is the negative-frequency part of the field and $\chi(z)=4\varepsilon_0\mathcal{A}d(z)$ is the coupling coefficient \cite{Horoshko24} with $d(z)$ the second-order nonlinear susceptibility of the crystal. In a periodically poled crystal, $d(z)$ changes sign every distance of $\Lambda/2$, where $\Lambda$ is the poling period. i.e., represents a meander function. This function can be decomposed into Fourier series, where only the term of order $1$ affects the phase matching \cite{BoydBook,Horoshko17}. Thus, we write $d(z)\approx(2/\pi)d_\text{eff}\exp(2\pi iz/\Lambda)$. Substituting Eqs.~(\ref{FourierSignal}) and (\ref{G}) into Eq.~(\ref{evolution}), performing the integration, and applying the commutation relations, we obtain the spatial evolution equations similar to those introduced by Caves and Crouch \cite{Caves87}
\begin{equation}\label{evolution2}
\begin{split}
    \frac{d\epsilon_o(z,\Omega)}{dz} &=\gamma\int \alpha(\Omega+\Omega')\epsilon^{\dagger}_e(z,\Omega')e^{i\Delta(\Omega,\Omega')z}\frac{d\Omega'}{2\pi},\\
    \frac{d\epsilon_e(z,\Omega)}{dz} &=\gamma\int \alpha(\Omega+\Omega')\epsilon^{\dagger}_o(z,\Omega')e^{i\Delta(\Omega',\Omega)z}\frac{d\Omega'}{2\pi},
\end{split}
\end{equation}
where $\gamma=8\varepsilon_0\mathcal{A}d_\text{eff}\mathcal{E}_p\mathcal{E}_o\mathcal{E}_e/(\pi\hbar)$ is the new coupling constant and $\Delta(\Omega,\Omega')=k_p(\Omega+\Omega')-k_o(\Omega)-k_e(\Omega')+2\pi/\Lambda$ is the phase mismatch for the three interacting waves. 

The solution of Eq. (\ref{evolution2}) can be written as an integral Bogoliubov transformation of the vacuum fields at the crystal input. For writing this transformation in a compact form, we introduce the envelopes of the ordinary wave at the crystal input and output faces by the relations $\hat E_o^{(+)}(0,t) = i\mathcal{E}_o A_0(t)e^{-i\omega_ot}$ and $\hat E_o^{(+)}(L,t) = i\mathcal{E}_o A_1(t)e^{i(k^0_oL-\omega_ot)}$ respectively, while those of the extraordinary wave at the same positions by the relations $\hat E_e^{(+)}(0,t) = i\mathcal{E}_e B_0(t)e^{-i\omega_et}$ and $\hat E_e^{(+)}(L,t) = i\mathcal{E}_e B_1(t)e^{i(k_e^0L-\omega_et)}$ respectively. Here, $k_\mu^0=k_\mu(0)$.  In these notations, the field transformation in the crystal is
\begin{eqnarray}\label{BogoliubovAfull}
A_1(t) &=& \int U_o(t,t')A_0(t')dt' + \int V_o(t,t')B_0^\dagger(t')dt',\\\label{BogoliubovBfull}
B_1(t) &=& \int U_e(t,t')B_0(t')dt' + \int V_e(t,t')A_0^\dagger(t')dt',
\end{eqnarray}
where the kernels $U_\mu(t,t')$ and $V_\mu(t,t')$ ($\mu=o,e$) satisfy the unitarity conditions
\begin{eqnarray}\label{unitarity1}
\int U_\mu(t,t')U_\mu^*(t'',t')dt' &-& \int V_\mu(t,t')V_\mu^*(t'',t')dt' \\\nonumber
&=& \delta(t-t''),\\\label{unitarity2}
\int U_o(t,t')V_e(t'',t')dt' &=& \int V_o(t,t')U_e(t'',t')dt',
\end{eqnarray}
guaranteeing the conservation of the commutation relations, $[A_i(t),A_i^\dagger(t')]=[B_i(t),B_i^\dagger(t')]=\delta(t-t')$, and $[A_i(t),B_i(t')]=0$, $i=0,1$. Analytical expressions for the Bogoliubov kernels $U_\mu(t,t')$ and $V_\mu(t,t')$ will be found in the low-gain regime in Sec.~\ref{sec:calculation}.
\\

\subsection{Correlation functions}

From Eqs. (\ref{BogoliubovAfull}) and (\ref{BogoliubovBfull}), we find that the mean field of the generated light is zero: $\langle A_1(t)\rangle = \langle B_1(t)\rangle = 0$. The field obtained from the vacuum by means of a Bogoliubov transformation is known to possess Gaussian statistics, which means that its higher-order moments are expressed via its second-order moments \cite{Mandel&Wolf}. Thus, all correlation functions can be obtained from the first-order ones: the normal autocorrelator and crosscorrelator $\langle \hat E_\mu^{(-)}(L,t)\hat E_\mu^{(+)}(L,t')\rangle$ and $\langle \hat E_o^{(-)}(L,t)\hat E_e^{(+)}(L,t')\rangle$, respectively, and the anomalous autocorrelator and crosscorrelator $\langle \hat E_\mu^{(+)}(L,t)\hat E_\mu^{(+)}(L,t')\rangle$ and $\langle \hat E_o^{(+)}(L,t)\hat E_e^{(+)}(L,t')\rangle$, respectively. For the field generated in a type-II PDC with vacuum at the input, the normal crosscorrelator and anomalous autocorrelator are identically zero, as can be easily found from Eqs. (\ref{BogoliubovAfull}) and (\ref{BogoliubovBfull}). The anomalous crosscorrelator, also known in the low-gain regime as biphoton amplitude, plays the central role in the description of entanglement between the generated photons and will be considered later. The only correlator to be calculated in this section is the normal autocorrelator of the ordinary wave,  $\langle \hat E_o^{(-)}(L,t)\hat E_o^{(+)}(L,t')\rangle$.

The second-order autocorrelation function of the ordinary wave is defined as \cite{Glauber63} 
\begin{widetext}
\begin{equation}\label{eq:socf1}
    g^{(2)}(\tau)=\frac{\left\langle \hat     E_{o}^{(-)}\left(L,t\right)\hat E_{o}^{(-)}(L,t+\tau)    \hat E_{o}^{(+)}(L,t+\tau)\hat E_{o}^{(+)}(L,t)\right\rangle}{\left\langle\hat E_{o}^{(-)}(L,t)\hat E_{o}^{(+)}(L,t) \right\rangle \left\langle \hat E_{o}^{(-)}(L,t+\tau)\hat E_{o}^{(+)}(L,t+\tau)\right\rangle}
\end{equation}
\end{widetext}
and is typically measured by a Hanbury Brown and Twiss setup shown in Fig. \ref{fig:scheme}, which involves the detection of single photons and a record of the time intervals between successive photon detections. 

The numerator of Eq. (\ref{eq:socf1}) is expressed via first-order correlators by the Gaussian moment factoring theorem \cite{Mandel&Wolf} (we omit the position $L$ for simplicity),
\begin{align}
    &\langle \hat E_{o}^{(-)}(t)\hat E_{o}^{(-)}(t+\tau)\hat E_{o}^{(+)}(t+\tau)\hat E_{o}^{(+)}(t)\rangle\nonumber \\
    &=\langle \hat E_{o}^{(-)}(t)\hat E_{o}^{(+)}(t)\rangle\langle \hat E_{o}^{(-)}(t+\tau)\hat E_{o}^{(+)}(t+\tau)\rangle\nonumber \\
    &+|\langle \hat E_{o}^{(-)}(t)\hat E_{o}^{(+)}(t+\tau)\rangle|^{2}+ |\langle \hat E_{o}^{(+)}(t)\hat E_{o}^{(+)}(t+\tau)\rangle|^{2}.\label{eq:gaussianfactor}
\end{align}
Upon discarding the last term, the anomalous autocorrelator being identically equal to zero, and substituting the rest into Eq. (\ref{eq:socf1}), we obtain
\begin{equation}\label{eq:socf3}
    g^{(2)}(\tau)=1+ |g^{(1)}(\tau)|^2,
\end{equation}
where
\begin{equation}\label{eq:normalized First order CF}
    g^{(1)}(\tau)=\frac{G^{(1)}(t,\tau)}{\left[I_o(t) I_o(t+\tau)\right]^{1/2}}.
\end{equation}
is the normalized first-order correlation function \cite{Glauber63}. Here, we have introduced the nonnormalized first-order correlation function $G^{(1)}(t,\tau)=\langle \hat E_{o}^{(-)}(L,t)\hat E_{o}^{(+)}(L,t+\tau)\rangle$ and the mean intensity of the ordinary wave $I_{o}(t)=G^{(1)}(t,0)$. Substituting Eq. (\ref{BogoliubovAfull}) into the definition of $G^{(1)}(t,\tau)$, we obtain
\begin{equation}\label{eq:focfbogo}
    G^{(1)}(t,\tau) = \mathcal{E}_o^2e^{-i\omega_o\tau}\int V_{o}^{*}(t,t')V_{o}(t+\tau,t')dt'.
\end{equation}
This function will be calculated analytically in the low-gain regime of PDC in Sec. \ref{sec:calculation}.

\subsection{Temporal imaging \label{sec:TI}}

Single picosecond-scale pulses of light generated by PDC in typical nonlinear crystals are impossible to resolve by modern photodetectors. Here, we study a possibility of stretching the field envelope of the ordinary wave by means of a temporal imaging system. A single-lens temporal imaging system consists of an input dispersive medium, a time lens, and an output dispersive medium, and it realizes the temporal stretching or compressing of a waveform similar to the manipulation of a wavefront by a single-lens imaging system in space \cite{Kolner94,Patera18}. 

We consider an optical wave with the carrier frequency $\omega_o$ propagating through a transparent medium of length $l$ and experiencing dispersion characterized by the dependence of its wave vector $k(\Omega)$ on the frequency $\omega=\omega_o+\Omega$. We decompose $k(\Omega)$ around the carrier frequency in powers of $\Omega=\omega-\omega_o$ and limit the Taylor series to the first three terms, $k(\Omega) \approx k^{0}+k'\Omega + \frac12k''\Omega^2$, where $k' = (d k/d \Omega)_{\omega_o}$ is the inverse group velocity, and $k'' = (d^2 k/d \Omega^2)_{\omega_o}$ is the group velocity dispersion of the medium at the carrier frequency $\omega_o$.

The spread of the initial waveform is governed by the second-order term, specifically through the group delay dispersion (GDD) denoted as $D=k''l$. The dispersive characteristics of the input and output media in the temporal imaging system are represented by their respective GDDs, $D_\mathrm{in}$ and $D_\mathrm{out}$. 

A time lens is a device imprinting a quadratic-in-time phase modulation on the transmitted waveform \cite{Telegin85,Kolner89}. Optical time lenses are engineered through electro-optical phase modulation (EOPM) \cite{Kolner94}, through nonlinear processes such as cross-phase modulation \cite{Mouradian00}, sum-frequency generation (SFG) \cite{Bennett00a, Bennett00b}, or four-wave mixing (FWM)\cite{Foster08,Meir23}, and through atomic-cloud-based quantum memory \cite{Mazelanik20,Mazelanik22,Niewelt23}. In any realization, the time lens is characterized by its focal GDD, $D_\mathrm{f}$, an ideal time lens providing a transformation $A_\text{out}(t)=A_\text{in}(t)\exp\left(it^2/2D_\mathrm{f}\right)$, where $A_\text{out}(t)$ is the field envelope in the group-delayed reference frame \cite{Kolner94,Patera18}. A realistic time lens can provide this modulation in a limited time window only, which is known as the temporal aperture of the time lens. In order for the above ideal transformation to be valid, it is necessary that the temporal object, upon stretching in the input dispersive medium, is shorter than the temporal aperture \cite{Srivastava23b}. A parametric lens, where the fields $A_\text{in}(t)$ and $A_\text{out}(t)$ have typically different carrier frequencies, should also have a frequency conversion efficiency close to unity within the temporal aperture \cite{Patera23}. In addition, the time lens should be synchronized with the PDC: the vertex of the parabola representing the time dependence of the modulating phase should coincide with the time of passage of the ordinary wave peak through the time lens \cite{Srivastava23}. When all these conditions and the temporal imaging condition
\begin{equation}\label{condition}
    \frac{1}{D_\mathrm{in}}+\frac{1}{D_\mathrm{out}}=\frac{1}{D_\mathrm{f}}
\end{equation}
are satisfied, the group-delayed output field envelope $A_2(t)$ defined via $\hat E^{(+)}_o(z_d,t) = i\mathcal{E}_o A_2(t-\tau_g) e^{ik_0z_d-i\omega_0t}$, where $z_d$ is the position after the temporal imaging system and $\tau_g$ is the group delay in all its elements, can be expressed via the input field envelope $A_1(t)$ in a rather simple way \cite{Patera18}
\begin{equation}\label{LensEq}
    A_2(t) = \frac{1}{\sqrt{M}}
    \exp\left(\frac{it^2}{2MD_\mathrm{f}}\right)A_1(t/M),
\end{equation}
where $M=-D_\mathrm{out}/D_\mathrm{in}$ is the magnification. As we can see, the output field intensity is a temporally scaled copy of the input field intensity, which is known as ``temporal imaging'' \cite{Kolner89,Kolner94}. We also see that the field amplitude, in addition to temporal scaling, has a chirp described by the term $e^{it^2/2MD_\mathrm{f}}$, which in spatial imaging corresponds to the wavefront curvature in the image plane. This chirp can be removed in a two-time-lens imaging system \cite{Srivastava23b}, but in the direct detection case considered here, it is irrelevant.

The nonnormalized first-order correlation function for the field after the temporal imaging system is $G^{(1)}_\text{im}(t,\tau)=\langle \hat E_{o}^{(-)}(z_d,t)\hat E_{o}^{(+)}(z_d,t+\tau)\rangle$. The modulus of this function is just a scaled and delayed version of the modulus of the function $G^{(1)}(t,\tau)$ analyzed in the previous section:
\begin{equation}\label{Gim}
   |G^{(1)}_\text{im}(t,\tau)| = \frac1{|M|}|G^{(1)}(t'/M,\tau/M)|,
\end{equation}
where $t'=t-\tau_g$ is the group-delayed time. This means that for sufficiently large magnification $|M|$, temporal variations of both the mean intensity $|G^{(1)}_\text{im}(t,0)|$ and the normalized second-order autocorrelation function $g^{(2)}_\text{im}(\tau)=g^{(2)}(\tau/M)$ can be brought to the time scale exceeding the resolution time of the photodetectors.

\section{Calculation of the second-order autocorrelation function \label{sec:calculation}}

In this section, we calculate the second-order autocorrelation function of one of the entangled beams, making three main approximations: approximation of the low-gain regime, where at most one pair of photons is generated per pump pulse, approximation of linear in frequency dispersion of the nonlinear crystal, and the Gaussian modeling of the phase-matching function. 

\subsection{Low-gain regime of PDC}

In the general case, the solution of Eq. (\ref{evolution}) represents a space-ordered exponent \cite{Quesada15} requiring a numerical treatment \cite{Christ13}, the analytical solution being known only for the case of a monochromatic pump \cite{Lipfert18}. However, in the low-gain regime of PDC, the probability of photon pair (biphoton) generation per pump pulse, $P_b$, is so small that the Bogoliubov kernels can be found perturbatively in the first non-vanishing order of $P_b$.

Equation~(\ref{evolution2}) can be solved perturbatively by substituting $\epsilon^{\dagger}_\mu(z,\Omega')\to\epsilon^{\dagger}_\mu(0,\Omega')$ under the integral. In this way, we obtain the fields at the output of a crystal of length $L$
\begin{equation}\label{CrystalSolution}
\begin{split}
    \epsilon_o(L,\Omega)&=\epsilon_o(0,\Omega) +\int    J(\Omega,\Omega')\epsilon^{\dagger}_e(0,\Omega') \frac{d\Omega'}{2\pi},\\  \epsilon_e(L,\Omega)&=\epsilon_e(0,\Omega) +\int   J(\Omega',\Omega)\epsilon^{\dagger}_o(0,\Omega') \frac{d\Omega'}{2\pi},
\end{split}
\end{equation}
where $J(\Omega,\Omega')=\gamma L\alpha(\Omega+\Omega')\Phi(\Omega,\Omega')$ is the joint spectral amplitude (JSA) of two generated photons and $\Phi(\Omega,\Omega')=e^{i\Delta(\Omega,\Omega')L/2}\sinc[\Delta(\Omega,\Omega')L/2]$ is the phase-matching function. 

A typical approximation in type-II PDC consists in linearizing the dispersion law: $k_\mu(\Omega)\approx k_\mu^0+k_\mu'\Omega$, where $k_\mu^0=k_\mu(0)$ and $k_\mu'$ is the derivative of $k_\mu$ with respect to $\Omega$ at $\Omega=0$, having the meaning of the inverse group velocity. In this approximation, the phase mismatch angle takes the form $\Delta(\Omega,\Omega')L/2\approx \tau_o\Omega+\tau_e\Omega'$, where $\tau_\mu=(k_p'-k_\mu')L/2$ and we assume that the poling period $\Lambda$ is chosen so that the condition of phase matching at degeneracy $k_p^0-k_o^0-k_e^0+2\pi/\Lambda=0$ is satisfied. The time $\tau_\mu$ has the meaning of the interval by which the peak of the corresponding subharmonic pulse advances the peak of the pump pulse at the crystal output: The peak of the subharmonic is generated when the pump pulse is in the center of the crystal, the travel times for the pump and subharmonic peaks through the second half of the crystal are $k_p'L/2$ and $k_\mu'L/2$, respectively; hence the meaning of $\tau_\mu$ as the difference of these times \cite{Srivastava23}. The JSA in this approximation reads 
\begin{eqnarray}\label{JSA}
J(\Omega,\Omega')&=&\gamma L\alpha_0e^{-\frac{(\Omega+\Omega')^2}{4\Omega_p^2} +i(\tau_o\Omega+\tau_e\Omega')}\\\nonumber
&&\times\sinc(\tau_o\Omega+\tau_e\Omega').
\end{eqnarray}

Substituting the solutions, Eq. (\ref{CrystalSolution}), into Eq. (\ref{FourierSignal}), we obtain the field transformation from the input of the crystal to its output. To write this transformation in compact form, we define the group-delayed envelopes: $\hat E_o^{(+)}(L,t)= i\mathcal{E}_oA_{1d}(t-k_o'L-\tau_o) e^{ik_o^0L-i\omega_ot}$ and $\hat E_e^{(+)}(L,t)= i\mathcal{E}_e B_{1d}(t-k_e'L-\tau_e) e^{ik_e^0L-i\omega_et}$. In this notation, the field transformation in the crystal has the form 
\begin{eqnarray}\label{BogoliubovA}
A_{1d}(t) &=& A_{0d}(t)  + \int \tilde J_0(t,t') B_{0d}^\dagger(t')dt',\\\label{BogoliubovB}
B_{1d}(t) &=& B_{0d}(t) + \int \tilde J_0(t',t) A_{0d}^\dagger(t')dt',
\end{eqnarray}
where $A_{0d}(t)=A_{0}(t+\tau_o)$ and $B_{0d}(t)=B_{0}(t+\tau_e)$ are delayed vacuum fields, while the zero-centered joint temporal amplitude (JTA) of two generated photons $\tilde J_0(t,t')$ is the double Fourier transform of their phase-shifted JSA $J_0(\Omega,\Omega')=J(\Omega,\Omega')e^{-i(\tau_o\Omega+\tau_e\Omega')}$:
\begin{equation}\label{JtildeDef}
\tilde J_0(t,t') = \iint J_0(\Omega,\Omega') e^{-i\Omega t-i\Omega't'}\frac{d\Omega d\Omega'}{(2\pi)^2}.
\end{equation}

The first-order autocorrelation function of the delayed ordinary wave is 
\begin{eqnarray}\label{Gddef}
G^{(1)}_d(t,\tau) &=& \mathcal{E}_o^2\langle A_{1d}^\dagger(t)A_{1d}(t+\tau)\rangle \\\nonumber
&=& \mathcal{E}_o^2\int \tilde J_0^*(t,t') \tilde J_0(t+\tau,t') dt'.
\end{eqnarray}
The function introduced in Sec. \ref{sec:Framework} differs from it by a delay and a phase: $G^{(1)}(t,\tau)=G^{(1)}_d(t-k_o'L-\tau_o,\tau)e^{-i\omega_o\tau}$. Thus, in the following we will calculate the delayed first-order autocorrelation function, which carries all necessary information on the field coherence. The correspondingly delayed intensity is denoted as $I_d(t)=G^{(1)}_d(t,0)$.

\subsection{Gaussian modeling}
Both JSA and JTA can be significantly simplified by approximating them with double Gaussian functions. For this purpose, we replace the $\sinc(x)$ function in Eq. (\ref{JSA}) with the Gaussian function $e^{-x^2/2\sigma_s^2}$ having the same width at half-maximum for $\sigma_s=1.61$ \cite{Grice01,Horoshko19}. Upon this replacement, the JSA becomes a double Gaussian function
\begin{equation}\label{JSAgauss}
J_0(\Omega,\Omega') = \gamma L\alpha_0\exp\left[-\frac{(\Omega+\Omega')^2}{4\Omega_p^2}-\frac{(\tau_o\Omega+\tau_e\Omega')^2}{2\sigma_s^2}\right].
\end{equation}
Substituting it into Eq. (\ref{JtildeDef}) and applying the multidimensional Gaussian integration technique \cite{Srivastava23,Srivastava23b}, we obtain the JTA also in the form of a double Gaussian function
\begin{equation}\label{JtildeGauss}
\tilde J_0(t,t') 
= J_1 e^{-M_{11}t^2-M_{22}{t'}^2-2M_{12}tt'},
\end{equation}
where $J_1=\gamma L\alpha_0\Omega_p^2/\pi|T_o-T_e|$, $T_o$ and $T_e$ are dimensionless advance times of the respective waves defined as $T_\mu=\sqrt{2}\Omega_p\tau_\mu/\sigma_s$, and $M_{ij}$ are elements of the matrix 
\begin{equation}\label{M}
    \mathbf{M} = \frac{\Omega_p^2}{(T_o-T_e)^2}\left(
    \begin{array}{cc}
        1+T_e^2 & -1-T_oT_e \\
        -1-T_oT_e & 1+T_o^2
    \end{array}
    \right).
\end{equation}

Substituting Eq. (\ref{JtildeGauss}) in Eq. (\ref{Gddef}) and performing the integration (see Appendix \ref{sec:appendixa}), we obtain
\begin{equation}\label{Gd}
    G^{(1)}_d(t,\tau)=\left[I_d(t)I_d(t+\tau)\right]^{1/2}\ e^{-\tau^{2}/2\Delta\tau_o^{2}},
\end{equation}
where 
\begin{equation}\label{Imu}
I_d(t) = \frac{\mathcal{E}_o^2P_b}{\sqrt{2\pi}\Delta t_o}e^{-t^2/2\Delta t_o^2}
\end{equation}
is the mean intensity of the ordinary wave, 
\begin{equation}\label{Pb}
P_b = \iint \left|\tilde J_0(t,t')\right|^2 dtdt' = \frac{(\gamma L\alpha_0\Omega_p)^2}{2\pi|T_e-T_o|}  
\end{equation}
is the probability of biphoton generation per pump pulse,
\begin{equation}\label{dt}
\Delta t_o = \frac{\sqrt{1+T_o^2}}{2\Omega_p}
\end{equation}
is the standard deviation of the mean intensity, and 
\begin{equation}\label{dtau}
    \Delta\tau_{o}=\frac{|T_o-T_e|\sqrt{1+T_{o}^{2}}}{\Omega_{p}|1+T_{o}T_{e}|}
\end{equation}
is the coherence time.

The correlation function (\ref{Gd}) describes a field emitted by a Gaussian Schell-model source \cite{Mandel&Wolf}. In the spatial domain, the ratio between the coherence length and the standard deviation of the transverse intensity distribution is known as the ``degree of global coherence'' \cite{Mandel&Wolf} of a partially coherent light beam. In a similar way, we introduce the (temporal) degree of global coherence as 
\begin{equation}\label{Q}
    C = \frac{\Delta\tau_{o}}{\Delta t_o} = 2\left|\frac{T_o-T_e}{1+T_oT_e}\right|,
\end{equation}
which tends to zero for a completely incoherent field and to infinity for a fully coherent one. This quantity can be inferred from the experimental data of temporally magnified photodetection described in Sec. \ref{sec:TI}. First, the distribution of arrival times on every detector in Fig. \ref{fig:scheme} is determined by the temporal profile of the magnified ordinary wave, which is given by Eq. (\ref{Gim}) as 
\begin{equation}\label{Iim}
   I_\text{im}(t) = |G^{(1)}_\text{im}(t,0)| = \frac1{|M|}I_d(t/M),
\end{equation}
and, taking Eq. (\ref{Imu}) into account, we see that the standard deviation of photon arrival times is $\Delta t_M = M\Delta t_o$. Second, as shown in Sec. \ref{sec:TI}, the normalized second-order autocorrelation function of the magnified wave is $g^{(2)}_\text{im}(\tau)=g^{(2)}(\tau/M)$.  Therefore, we have from Eqs. (\ref{eq:socf3}), (\ref{eq:normalized First order CF}), and (\ref{Gd}) that
\begin{equation}\label{gim}
   g^{(2)}_\text{im}(\tau) = 1 + e^{-\tau^{2}/M^2\Delta\tau_o^{2}},
\end{equation}
and a Gaussian fit of the function $g_\text{im}^{(2)}(\tau)-1$ will give a standard deviation $\Delta \tau_M = M\Delta \tau_o/\sqrt{2}$. As a result, the degree of global coherence can be found as $C=\sqrt{2}\Delta \tau_M/\Delta t_M$.

The described method of measuring the degree of global coherence is insensitive to the loss which may occur between the crystal and the detectors, including the non-unit quantum efficiency of the latter. Indeed, linear loss does not affect the normalized temporal profile of the optical pulse, and therefore does not change $\Delta t_M$. On the other hand, the normalized second-order autocorrelation function, and therefore its width $\Delta\tau_M$, are well-known to be loss-insensitive. Even a very high loss in the temporal imaging system and photodetectors will affect the time necessary for gathering a large enough statistics but not the quality of the obtained result.

\subsection{Schmidt decomposition}

A double-Gaussian JTA can be represented in a form of Schmidt decomposition by means of Mehler's formula for Hermite polynomials (see Appendix~\ref{sec:appendixb}) \cite{Horoshko23b} 
\begin{equation}\label{Schmidt}
\tilde J_0(t,t') 
= \sqrt{P_b}\sum\limits_{n=0}^{\infty}\sqrt{\lambda_n}\psi_n(t)\varphi_n(t'),
\end{equation}
where $\psi_n(t)$ and $\varphi_n(t')$ are the Schmidt modal functions of the ordinary and extraordinary photons respectively, defined via the Hermite-Gauss functions $h_n(x)$ as
\begin{eqnarray}\label{psin}
\psi_n(t) &=& \frac1{\sqrt{\tau_1}}h_n(t/\tau_1),\\\label{phin}
\varphi_n(t) &=& \frac1{\sqrt{\tau_2}}h_n(t/\tau_2),
\end{eqnarray}
$\lambda_n$ are the Schmidt coefficients
\begin{equation}\label{lambda}
\lambda_n=\frac2{K+1}\left(\frac{K-1}{K+1}\right)^n  
\end{equation}
normalized to unity, $\sum_n\lambda_n=1$, and 
\begin{equation}\label{K}
K=\frac1{\sum_n\lambda_n^2} = \frac{\sqrt{(1+T_o^2)(1+T_e^2)}}{|T_o-T_e|}  
\end{equation}
is the Schmidt number. The Schmidt number $K$ shows the effective number of entangled modes for each photon and as such is a measure of the degree of entanglement \cite{Law04,Horoshko12,Gatti12}. The temporal scales of the Schmidt modes are 
\begin{equation}\label{tau12}
\tau_{1,2} = \frac{\sqrt{|T_o-T_e|}}{\sqrt{2}\Omega_p}\left(\frac{1+T_{o,e}^2}{1+T_{e,o}^2}\right)^\frac14.
\end{equation}

Substituting the decomposition (\ref{Schmidt}) into Eq. (\ref{Gddef}) and using the completeness of modal functions, we obtain
\begin{equation}\label{Mercer}
G^{(1)}_d(t,\tau)= \mathcal{E}_o^2 P_b\sum\limits_{n=0}^{\infty}\lambda_n \psi_n^*(t)\psi_n(t+\tau),
\end{equation}
which is known as the Mercer expansion in coherence modes \cite{Mandel&Wolf}. We see that the Schmidt modes of two entangled waves are the coherence modes of each wave. From Eqs. (\ref{Q}) and (\ref{K}) we obtain 
\begin{equation}\label{KQ}
K^2 = 1+\frac4{C^2},  
\end{equation}
which relates the measure of entanglement $K$ to the measure of coherence $C$. Equation (\ref{KQ}) has a clear physical meaning: The higher the coherence, the lower the entanglement, with a single-mode regime ($K=1$) for a fully coherent ordinary wave ($C=\infty$). In addition, this equation allows one to experimentally determine the number of modes by measuring $C$, as described in the preceding section.  

It should be noted that the Mercer expansion of the first-order correlation function of one of the two entangled beams was used to determine the spatial Schmidt modes of that beam \cite{Sharapova20,Averchenko20,Kopylov25}. A similar technique was applied to determine the spatial Schmidt modes of a pseudo-thermal light \cite{Bobrov13}. However, the relationship between single-beam coherence and cross-beam entanglement measures was not previously analyzed, to the best of our knowledge. 

In addition to what was said in the preceding section about the influence of loss, we note that the loss occurring inside the nonlinear crystal affects the modal structure of the field and requires a more complicated model \cite{Kopylov25} and a separate study. This kind of loss can typically be neglected in bulk crystals, but may be significant in relatively long nonlinear waveguides.

\subsection{Symmetric group velocity matching \label{sec:symm}}
We illustrate the general formalism by considering a crystal with $\tau_o=-\tau_e$, which is known as symmetric group velocity matching \cite{Keller97,Ansari18} or extended phase matching \cite{Giovannetti02}. This type of phase matching is attractive because it allows one to pass from the single-mode to a multi-mode regime simply by varying the pump spectral bandwidth \cite{Brecht15}. It was first engineered in a crystal of periodically poled potassium titanyl phosphate (ppKTP) \cite{Konig04}, which we consider here as an example. The pump propagates along the $X$ axis of this biaxial crystal and is polarized along the $Y$ axis, i.e. represents an ordinary wave. The two subharmonic waves are polarized along the $Y$ (ordinary wave) and $Z$ (extraordinary wave) axes of the crystal. The latter should not be confused with the $z$ axis of the reference frame, the direction of pump propagation.

The refractive indices for the $Y$ and $Z$ polarized waves are obtained from the Sellmeier equations for KTP \cite{Konig04} and, for a crystal of length $L=40$ mm pumped at wavelength $\lambda_p=2\pi c/\omega_p=791.5$ nm, we obtain $\tau_o=-\tau_e=2.95$ ps. The poling period required to reach the quasi-phase matching at degeneracy is $\Lambda=47.6$ $\mu$m. Substituting $\tau_o=-\tau_e>0$ into Eq. (\ref{K}), we obtain
\begin{equation}\label{Ksymm}
K= \frac12\left(\frac1{T_o}+T_o\right) = \frac12\left(\frac{\tau_p}{\delta_s\tau_o}+\frac{\delta_s\tau_o}{\tau_p}\right),
\end{equation}
where $\delta_s=2\sqrt{\ln2}/\sigma_s\approx1.03$. The minimal value of $K=1$ is reached at the pump pulse duration $\tau_p=\delta_s\tau_o\approx3$ ps, which corresponds to $\Omega_p = 2\pi\times 62$ GHz or FWHM spectral bandwidth $\Delta\lambda=0.3$ nm. In this case the photons are disentangled, each occupying just one Gaussian temporal mode, as required for some applications, for example, the heralded generation of modally pure single photons \cite{Mosley08}. 

\begin{figure}[ht!]
\centering
\includegraphics[width=0.99\columnwidth]{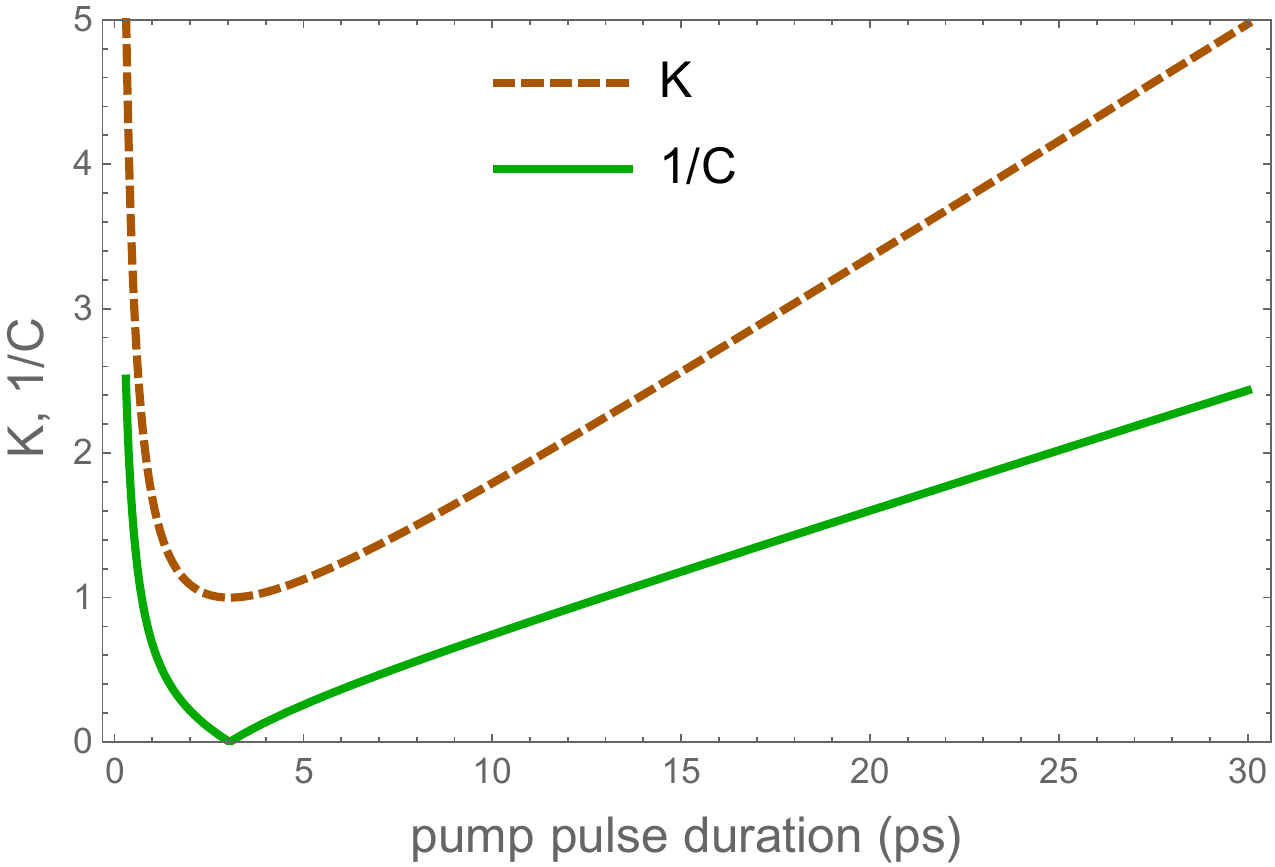}
\caption{Schmidt number and inverse degree of global coherence of the ordinary wave generated by PDC in a $40$-mm-long ppKTP crystal pumped at $\lambda_p = 791.5$ nm as functions of the pump pulse duration. \label{fig:InverseDGC}}
\end{figure}

Substituting $\tau_o=-\tau_e$ into Eq. (\ref{Q}), we obtain
\begin{equation}\label{Qsymm}
\frac1C = \frac14\left|\frac1{T_o} - T_o\right|= \frac14\left|\frac{\tau_p}{\delta_s\tau_o} -\frac{\delta_s\tau_o}{\tau_p} \right|
\end{equation}
showing that the inverse degree of global coherence tends to zero at the pump pulse duration $\tau_p=\delta_s\tau_o$. Figure \ref{fig:InverseDGC} shows this dependence together with the dependence of the Schmidt number, Eq. (\ref{Ksymm}).

\begin{figure*}[!ht]
\centering
\includegraphics[width=0.49\textwidth]{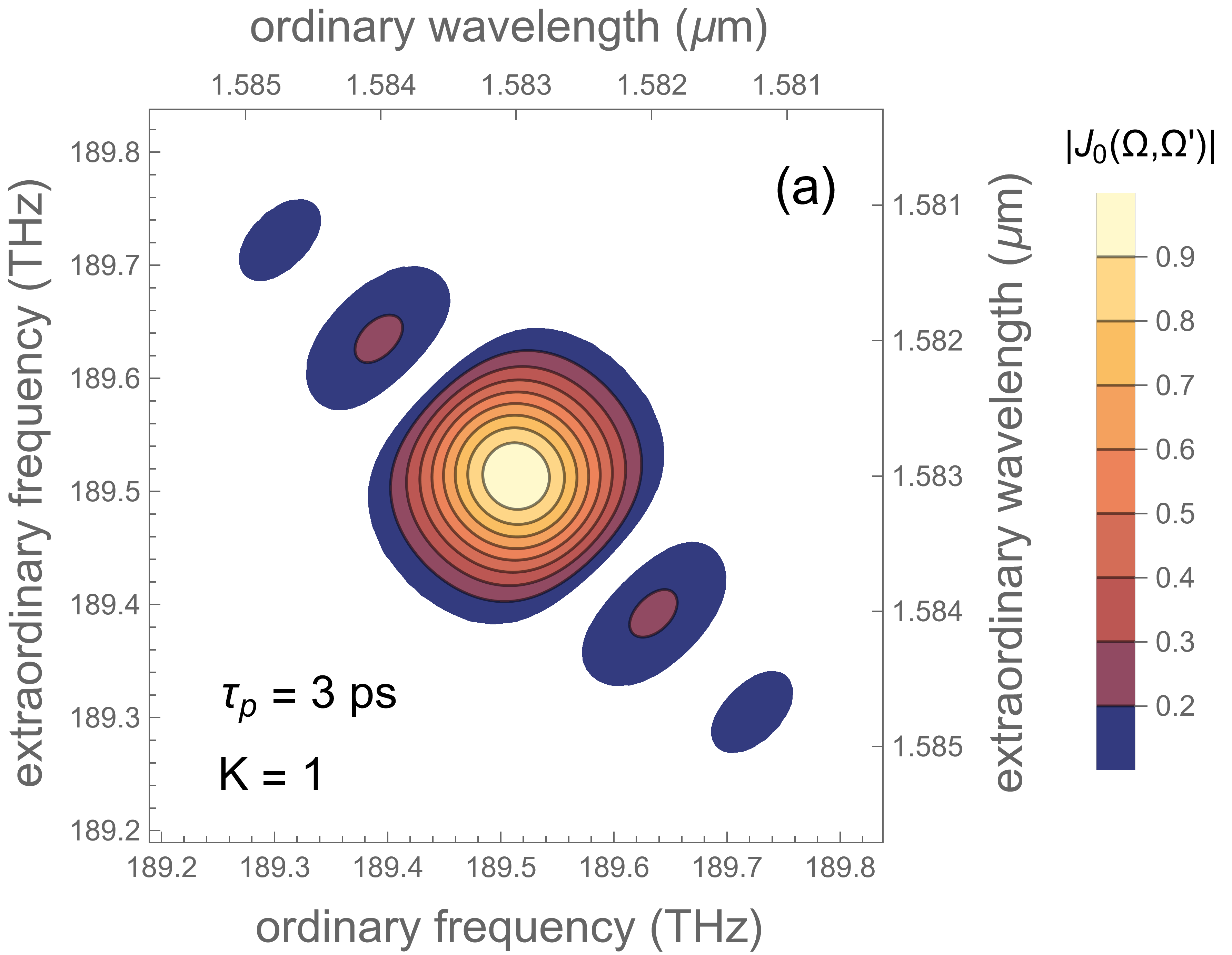}
\includegraphics[width=0.49\textwidth]{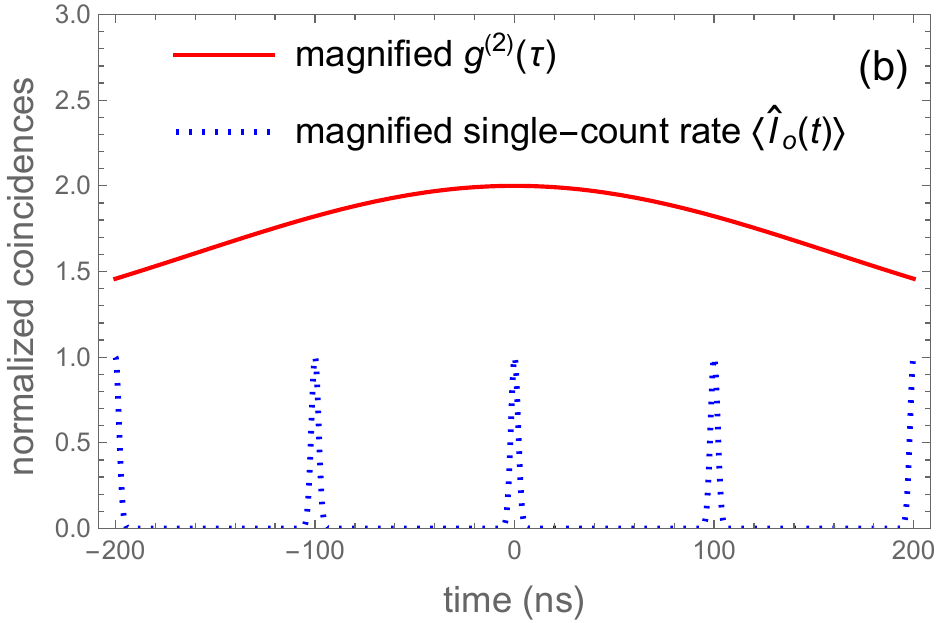}
\includegraphics[width=0.49\textwidth]{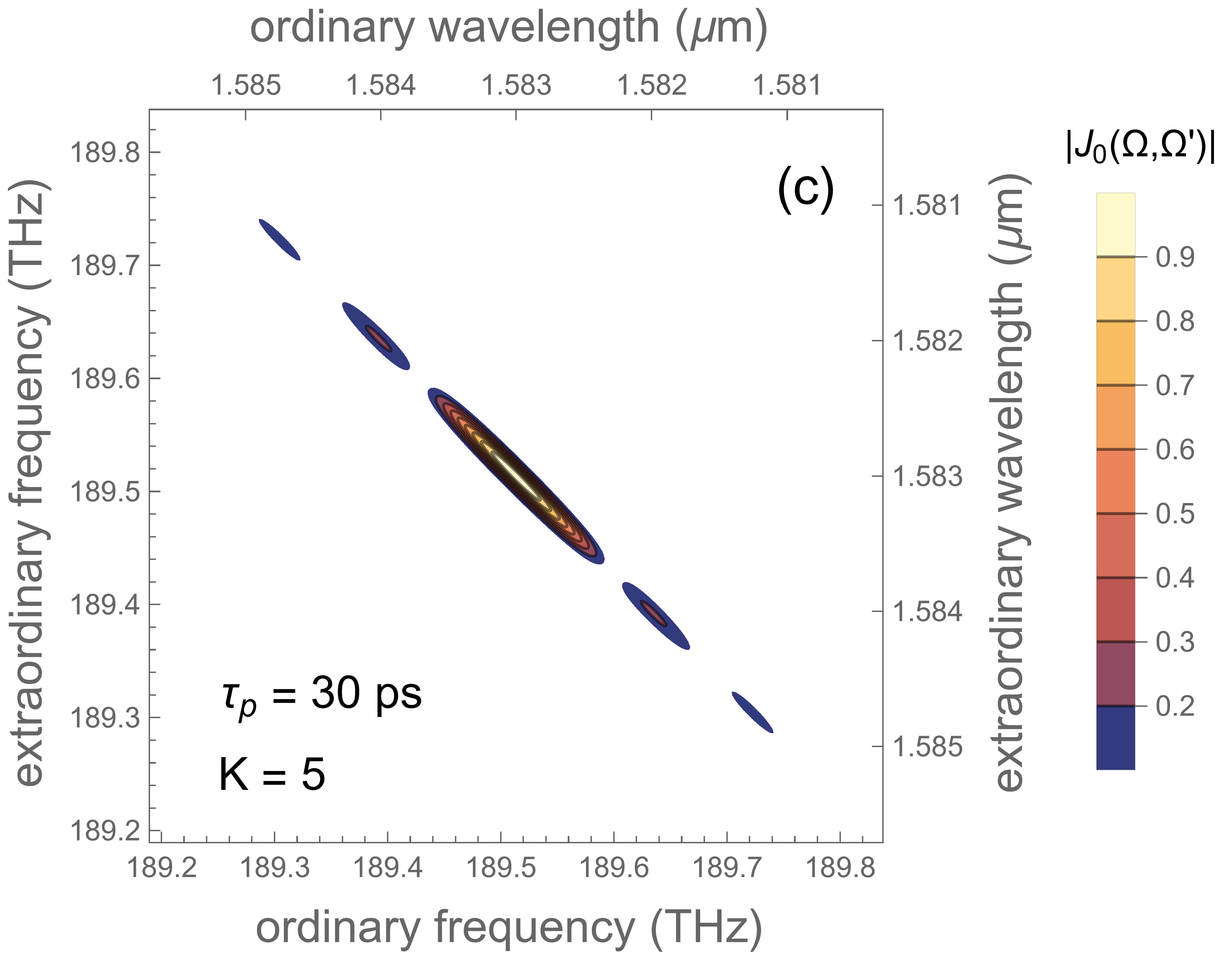}
\includegraphics[width=0.49\textwidth]{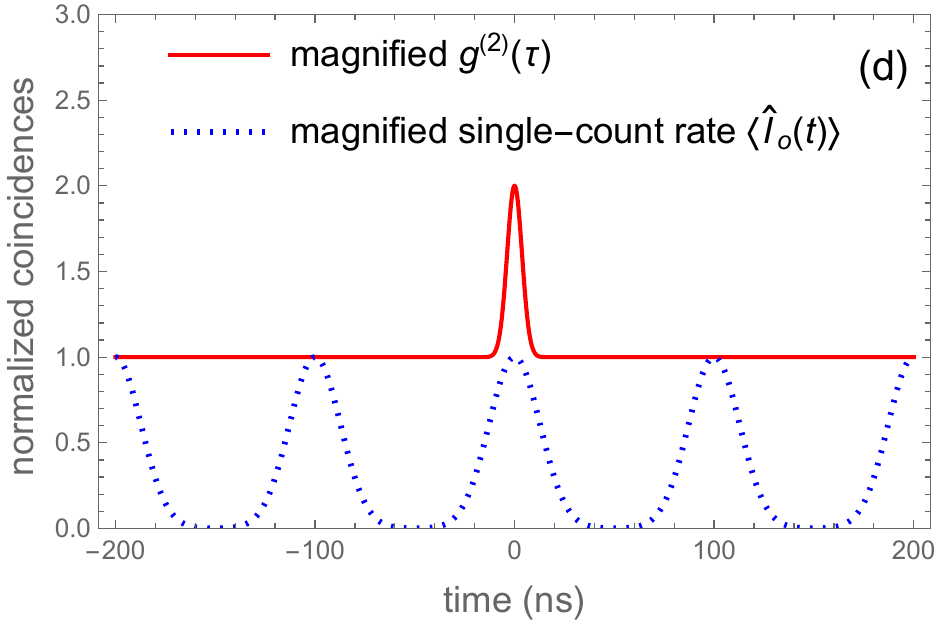}
\caption{(a,c) Normalized JSA and (b,d) temporally magnified $g^{(2)}(\tau)$ function and normalized mean intensity (single-count rate) of the ordinary wave generated by PDC in a $40$-mm-long ppKTP crystal pumped at $\lambda_p = 791.5$ nm by pulses of duration (a,b) 3 ps resulting in a single-mode regime, or (c,d) 30 ps resulting in a multi-mode regime. In the single-mode regime, the coherence time is much larger than the pulse duration. In the multi-mode regime, it is less than the pulse duration. Temporal magnification is $M=1000$, repetition rate is 10 MHz. \label{fig:KTP}}
\end{figure*}

A more detailed picture is provided by Fig. \ref{fig:KTP} showing the JSA of the photons and their normalized correlation function and mean intensity. For comparison, we consider the case of $\tau_p=3$ ps and also the case of a ten-times longer pump pulse with $\tau_p=30$ ps, which corresponds to $\Omega_p = 2\pi\times 6.2$ GHz or FWHM spectral bandwidth $\Delta\lambda=0.03$ nm. In the case of $\tau_p=3$ ps, we have $T_o\approx1$ and $K\approx1$. We find $\Delta t_o=1.8$ ps and $\Delta \tau_o=226$ ps, wherefrom $C=125$, which corresponds to Eq. (\ref{KQ}) with $K\approx1$. In the case of $\tau_p=30$ ps, we have $T_o\approx0.1$ and obtain $K\approx5$ from Eq. (\ref{Ksymm}). We also find $\Delta t_o=13$ ps and $\Delta \tau_o=5.3$ ps from Eqs. (\ref{dt}) and (\ref{dtau}), and obtain $C=0.41$. These values of $K$ and $C$ are related by Eq. (\ref{KQ}). The same values of $K$ and $C$ appear for pulses 10 times shorter than in the single-mode case, that is, for $\tau_p=0.3$ ps.

In subfigures (b) and (d) of Fig. \ref{fig:KTP}, we show the normalized single-count rate on both detectors as a function of the absolute time and the magnified $g^{(2)}(\tau)$ function as a function of the time difference. We see that, in a regime close to the single-mode one, the coherence time may significantly exceed the pulse duration. In practice, measuring a long coherence time similar to that shown in Fig. \ref{fig:KTP}(b) requires two conditions: (i) a high number of detection events $N$ and (ii) a low repetition rate $R$. Indeed, the standard deviation of the magnified function $g^{(2)}(\tau)-1$ is $M\Delta\tau_o/\sqrt{2}=MC\Delta t_o/\sqrt{2}$. Since photodetection events have a Gaussian distribution in time with the standard deviation $M\Delta t_o$, we find that the average number of events outside the area $\pm M\Delta\tau_o/\sqrt{2}$ around the peak is $N_\text{out}=N\erfc(C/\sqrt{2})$. In order for the fit of the dependence of the magnified function $g^{(2)}(\tau)-1$ within its entire $\sigma$ area to be efficient, we expect $N_\text{out}\gg1$. Thus, for measuring $C=4.9$, one needs to satisfy the condition $N\gg10^6$, which is rather stringent. Higher values of $C$ require a large unpractical number of detection events. Another limitation concerns the repetition rate. Indeed, the pairs of photodetection events used to determine the $g^{(2)}(\tau)$ function should come from the same pump pulse, which requires the repetition period to be greater than the two standard deviations of the magnified function $g^{(2)}(\tau)-1$, i.e., $R\le(\sqrt{2}MC\Delta t_o)^{-1}$. The necessity of having a low repetition rate and a high number of detection events at the same time makes measuring high values of $C$ extremely challenging. However, even measurable values $0\le C\le 5$ provide deep insight into the modal structure of generated photons. We note that in the case of high temporal coherence, the traditional method of determining the number of modes by the time-integrated function $g^{(2)}_\text{int}$ provides excellent results \cite{Christ11}, as mentioned in the Introduction. 

Concluding this section, we discuss how the introduced quantities are obtained in an experiment. The intensity autocorrelation function is measured by selecting two intervals of width $\delta t$ around time $t_p+t$ on detector $D_1$ (interval 1) and around time $t_p+t+\tau$ on detector $D_2$ (interval 2), where $t_p$ is the time of the peak of the closest pump pulse. Denote by $N_1$ and $N_2$ the total numbers of photocounts in intervals 1 and 2, respectively, and by $N_{12}$ the number of coincidences in intervals 1 and 2. 

\begin{figure*}[ht!]
\centering
\includegraphics[width=18cm]{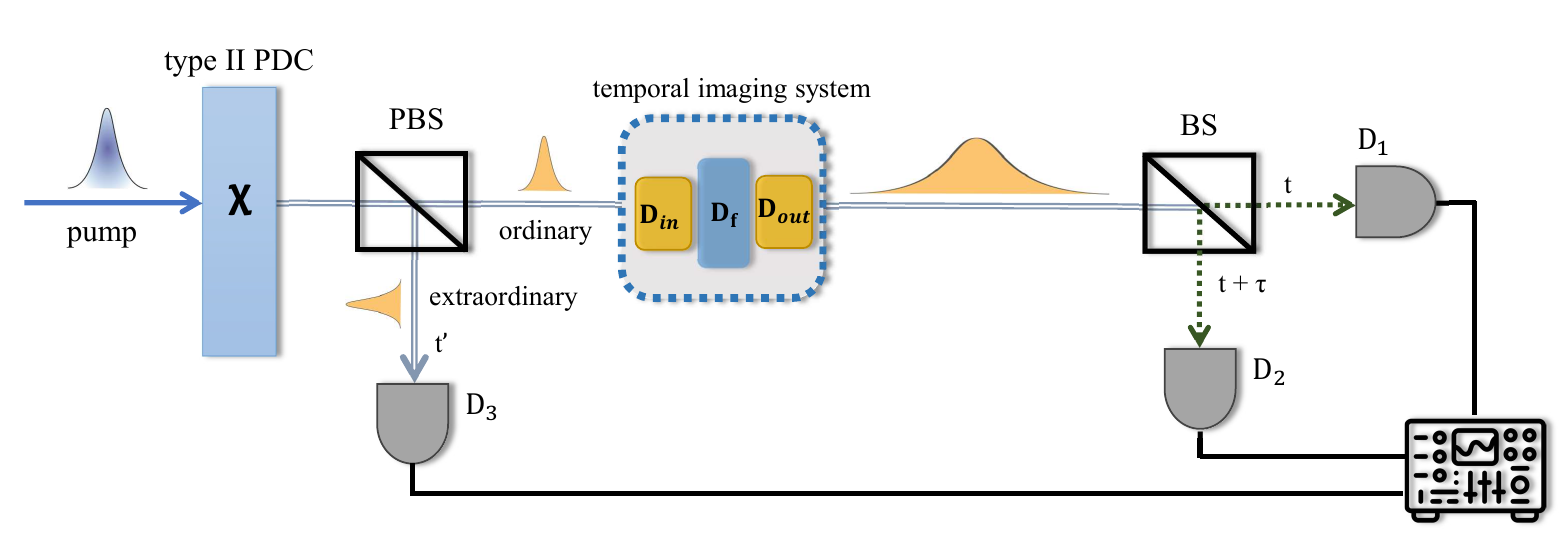}
\caption{Scheme similar to Fig. \ref{fig:scheme} to measure the time-resolved heralded second-order autocorrelation function. The only difference is the measurement of the extraordinary wave by a binary bucket detector $D_3$, which resolves neither the time of arrival nor the number of arriving photons. A click of this detector creates a heralding event and the photocount statistics of detectors $D_1$ and $D_2$ are conditioned on this event.\label{fig:heraldedscheme}}
\end{figure*}

We introduce delayed intensities (in photon flux units) of the ordinary and extraordinary waves, $\hat I_o(t)=A_{1d}^\dagger(t)A_{1d}(t)$ and $\hat I_e(t')=B_{1d}^\dagger(t')B_{1d}(t')$, respectively. Since $\hat I_o(t)$ is the operator of the photon flux of the ordinary wave, its average gives the single-count rate. Therefore, the probability of observing a photon in interval 1 is $P_1=F_o\langle \hat I_o(t)\rangle$, where $F_o=\frac12\delta t$, the factor $\frac12$ being the transmittance of the beam splitter in the ordinary arm. Similarly, the probability of observing a photon in interval 2 is $P_2=F_o\langle \hat I_o(t+\tau)\rangle$, and the probability of coincidence in these two intervals is $P_{12}=F_o^2\langle : \hat I_o(t)\hat I_o(t+\tau):\rangle$. To obtain the number of events, we need to multiply its probability by the total number $N$ of pump pulses: $N_1=P_1N$, $N_2=P_2N$, and $N_{12}=P_{12}N$. As a result, Eq. (\ref{eq:socf1}) can be rewritten as
\begin{equation}\label{g2exp}
g^{(2)}(\tau) = \frac{P_{12}}{P_1P_2} = \frac{N_{12}N}{N_1N_2},   
\end{equation}
and, in the ideal case considered here, this ratio should not depend on the absolute time $t$, but only on the time difference $\tau$.

\section{Heralded second-order autocorrelation function \label{sec:Heralded}}

In this section, we consider a more complicated scenario, where the extraordinary photon is not discarded, as above, but detected on detector $D_3$ (see Fig.~\ref{fig:heraldedscheme}) without photon-number nor temporal resolution (the so-called ``binary bucket'' detector), which produces a ``heralding event'' \cite{Mosley08}. Since the time of arrival of this photon remains unknown, we need to integrate the corresponding probability density throughout the pump repetition period $\Pi_t$ that contains the time $t$. 

The ``heralded'' autocorrelation function of the ordinary wave can be defined by extending Eq. (\ref{g2exp}) for triple coincidences. To this end, we define interval $h$ as the entire period $\Pi_t$ on detector $D_3$ and denote by $N_h$ the number of heralding events. We denote also by $N_{12h}$ the number of triple coincidences in intervals 1, 2 and $h$, and by $N_{ih}$ the number of double coincidences in intervals $i=1,2$ and $h$. The coincidence probability $P_{12}=N_{12}/N$ should be replaced in the heralded case by the conditional coincidence probability $P_{12|h}=N_{12h}/N_h$, which counts the statistical frequency of coincidences in the periods where a heralding event occurs. Here we imply that intervals 1 and $h$ belong to the same pump period, while interval 2 may belong to the same or different pump period.

We see two possible ways of modifying the denominator of Eq. (\ref{g2exp}) in the heralded case. First, we can replace the probability $P_1=N_1/N$ by the conditional probability $P_{1|h}=N_{1h}/N_h$, but leave $P_2$ unchanged. In this way, we introduce the single-heralded autocorrelation function,
\begin{equation}\label{g2shexp}
g^{(2)}_\text{sh}(\tau) = \frac{P_{12|h}}{P_{1|h}P_2} = \frac{N_{12h}N}{N_{1h}N_2} ,   
\end{equation}
which is an extension of Eq. (\ref{g2exp}) with the assumption that only the click of detector $D_1$ is heralded, while the click of detector $D_2$ is produced by an unheralded photon from another photon pair generated in the same or different period.

Second,  we can replace both probabilities $P_1$ and $P_2$ by their conditional variants $P_{i|h}=N_{ih}/N_h$ and introduce the double-heralded autocorrelation function,
\begin{equation}\label{g2dhexp}
g^{(2)}_\text{dh}(\tau) = \frac{P_{12|h}}{P_{1|h}P_{2|h}} = \frac{N_{12h}N_h}{N_{1h}N_{2h}} ,   
\end{equation}
which is an extension of Eq. (\ref{g2exp}) with the assumption that both clicks on detectors $D_1$ and $D_2$ are heralded by the same photon detection on detector $D_3$. This function represents a time-resolved version of the ``anticorrelation parameter'' introduced in Ref.~\cite{Grangier86}. It was shown that its value below 1 is a witness of light nonclassicality \cite{Grangier86,URen05}, which is also discussed in Refs. \cite{Bocquillon09,Bettelli10}.

The coincidence probabilities can be rather easily calculated from the analytical model of PDC presented in the previous section. The number of heralding events is obtained by calculating the probability of photodetection on detector $D_3$ at time $t'$ with a subsequent integration with respect to the latter:
\begin{equation}\label{Nh}
N_h = N\langle  \int_{\Pi_t} \hat I_e(t')dt'\rangle =  NP_b,   
\end{equation}
where the average was calculated from Eqs.~(\ref{BogoliubovA}) and (\ref{BogoliubovB}).

The number of double coincidences between detectors $D_1$ and $D_3$ (in the same pump period) is given in a similar way by
\begin{eqnarray}\label{N1h}
N_{1h} &=& NF_o\langle :\hat I_o(t) \int_{\Pi_t} \hat I_e(t')dt':\rangle \\\nonumber
&=& NF_o\langle \hat I_o (t)\rangle (1+P_b),   
\end{eqnarray}
where again we used Eqs.~(\ref{BogoliubovA}) and (\ref{BogoliubovB}) to find the average.

The number of double coincidences between detectors $D_2$ and $D_3$ in the same pump period is calculated as above. In different periods, the two fields are completely decorrelated. As a result, we have  
\begin{equation}\label{N2h}
N_{2h}=\left\{
\begin{array}{cc}
    NF_o\langle \hat I_o (t+\tau)\rangle (1+P_b), & \text{if } t+\tau\in \Pi_t, \\
    NF_o\langle \hat I_o (t+\tau)\rangle P_b, & \text{if } t+\tau\notin \Pi_t.
\end{array}
\right.   
\end{equation} 

\begin{figure}[ht!]
\centering
\includegraphics[width=0.99\columnwidth]{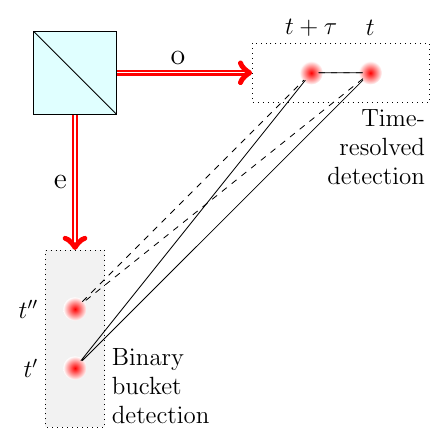}
\caption{Triple coincidences from four photons. Ordinary (o) and extraordinary (e) photons separated by a polarizing beam splitter, as shown in Fig.~\ref{fig:heraldedscheme}, are detected with or without temporal resolution. When a coincidence of two ordinary photons occurs at times $t$ and $t+\tau$, two extraordinary photons arrive to the detector of the extraordinary arm, which resolves neither the time of arrival nor the number of arriving photons. As a result, only the first triple coincidence (solid line triangle) is registered, but not the second one (dashed line triangle). \label{fig:4photons}}
\end{figure}

For calculating the probability of a triple coincidence, we need again to distinguish two cases: The detection on $D_2$ occurs in the same period as the detections on $D_1$ and $D_3$ or in a different period. In the first case, where $\tau$ is sufficiently small so that $t+\tau\in \Pi_t$, the number of triple coincidences in intervals 1, 2, and $h$ is 
\begin{equation}\label{N12h}
N_{12h}=\frac12 NF_o^2 \langle :\hat I_o(t)\hat I_o(t+\tau) \int_{\Pi_t} \hat I_e(t')dt':\rangle.   
\end{equation} 
where the factor $\frac12$ reflects the fact that the detector fixes only the first photon arriving to its surface, but not the second one, while the third-order intensity correlator gives the probability of all possible triples created by two ordinary and two extraordinary photons; see Fig.~\ref{fig:4photons}. When $\tau$ grows so high that $t+\tau$ falls into the next pump period, i.e., $t+\tau\notin \Pi_t$, the field $\hat I_o(t+\tau)$ is decorrelated from the other two fields and the number of triple coincidences in intervals 1, 2, and $h$ is 
\begin{equation}\label{N12hbis}
N_{12h}=NF_o^2 \langle :\hat I_o(t) \int_{\Pi_t} \hat I_e(t')dt':\rangle\langle \hat I_o(t+\tau)\rangle.   
\end{equation} 
Using again Eqs.~(\ref{BogoliubovA}) and (\ref{BogoliubovB}) to find the averages, we finally obtain
\begin{equation}\label{N12hter}
\frac{N_{12h}}{NF_o^2}=\left\{
\begin{array}{cc}
    \langle :\hat I_o(t)\hat I_o(t+\tau):\rangle (1+\frac12P_b), & \text{if } t+\tau\in \Pi_t, \\
    \langle \hat I_o(t)\rangle\langle \hat I_o(t+\tau)\rangle (1+P_b), & \text{if } t+\tau\notin \Pi_t.
\end{array}
\right.   
\end{equation} 

We see that in the limit of very low gain, where $P_b\ll1$, the number of heralded double coincidences on $D_1$ and $D_2$ is equal to the number of unheralded coincidences found in the preceding section. This is easy to understand, because, in the lossless model considered here, any pair of ordinary photons is accompanied by a pair of extraordinary photons, which trigger a heralding event with probability 1. For example, if we consider a realistic experiment with $N=10^9$ pump pulses and the probability of biphoton generation $P_b=10^{-2}$, we obtain the number of double ordinary photons generated in the same period as $P_b^2N=10^5$. In half of the cases, these photons will go to different detectors producing $N_{12}^\text{total}=50000$ coincidences, where the superscript ``total'' means a summation over all possible intervals of duration $\delta t$. In the lossless model considered here, every such coincidence will be accompanied by a click on detector $D_3$, which means that the number of triple coincidences is $N_{12h}^\text{total}=50000=N_{12}^\text{total}$. Similarly, we obtain the number of heralded single clicks $N_{1h}^\text{total}=P_bN=10^7=N_1^\text{total}$. It means that, in the limit of very low biphoton generation probability and in the absence of losses, heralding does not change the observed number of single or double clicks in the ordinary arm.

\begin{figure}[h!]
\centering
\includegraphics[width=0.99\columnwidth]{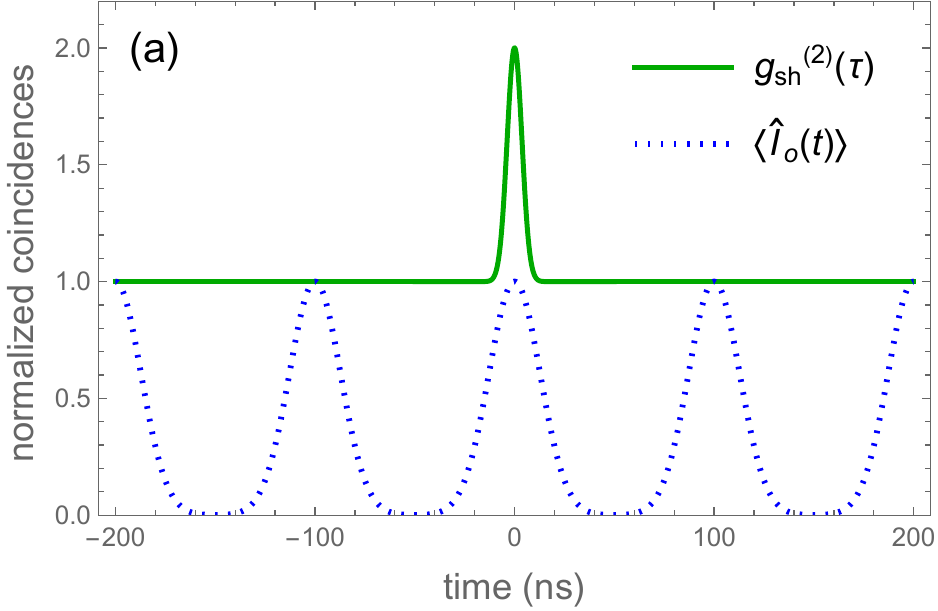}
\includegraphics[width=0.99\columnwidth]{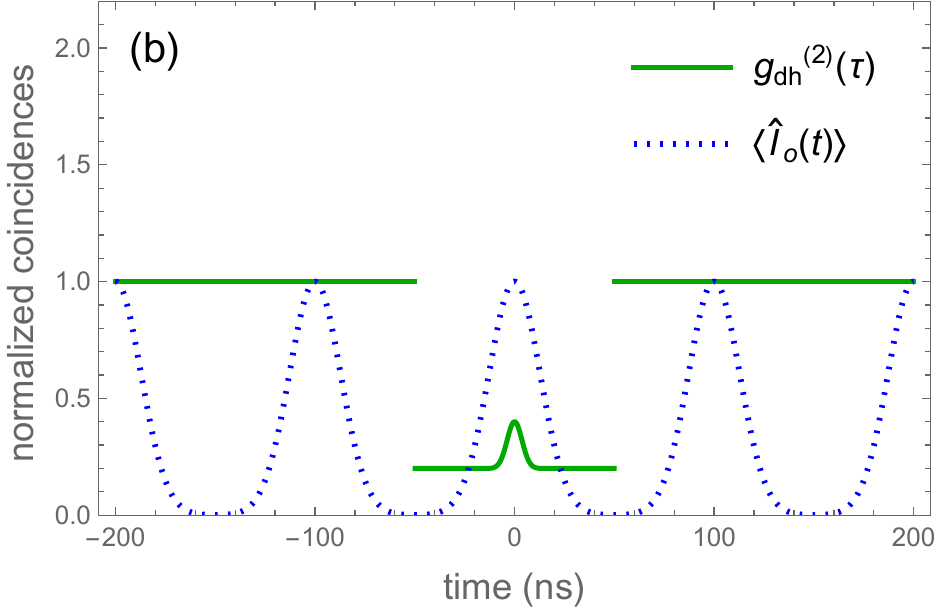}
\caption{Single- and double-heralded autocorrelation functions of the ordinary wave generated by PDC in a $40$-mm-long ppKTP crystal pumped at $\lambda_p = 791.5$ nm by pulses of duration 30 ps and measured with a temporal magnification $M=1000$. 
The solid curves correspond to Eqs. (\ref{sheralded}) and (\ref{dheralded}) in the limit of $P_b\ll1$, but, in subfigure (b), we set $P_b=0.2$ for the illustration purpose, because $g_\text{dh}^{(2)}(\tau)$ ranges from $P_b$ to $2P_b$ for $|\tau|$ below the half-repetition-period of the pump. \label{fig:heraldedG2}}
\end{figure}

Using the obtained expressions for the numbers of coincidences, we rewrite the single- and double-heralded autocorrelation functions, Eqs. (\ref{g2shexp}) and (\ref{g2dhexp}), in the case $t+\tau\in \Pi_t$ as 
\begin{eqnarray}\nonumber
g^{(2)}_\text{sh}(\tau)&=&\frac
    {\frac12\left\langle :\hat I_o(t)\hat I_o(t+\tau)\int_{\Pi_t} \hat I_e(t')dt':\right\rangle}
    {\left\langle :\hat I_o(t)\int_{\Pi_t} \hat I_e(t')dt':\right\rangle 
    \langle \hat I_o(t+\tau)\rangle} \\\label{sheralded}
    &=& g^{(2)}(\tau)\frac{1+\frac12P_b}{1+P_b}
\end{eqnarray}
and
\begin{eqnarray}\nonumber
    g_\text{dh}^{(2)}(\tau)&=&\frac{
    \frac12\langle : \hat I_{o}(t)I_{o}(t+\tau)\int \hat I_{e}(t')\mathrm{d}t': \rangle 
    \langle \int \hat I_{e}(t')\mathrm{d}t' \rangle}
    {\langle : \hat I_{o}(t)\int \hat I_{e}(t')\mathrm{d}t': \rangle
    \langle : \hat I_{o}(t+\tau)\int \hat I_{e}(t')\mathrm{d}t': \rangle}\\\label{dheralded}
    &=& g^{(2)}(\tau) \frac{(1+\frac12P_b)P_b}{(1+P_b)^2},
\end{eqnarray}
where all integrations are assumed to be over the period $\Pi_t$. For $t+\tau\notin \Pi_t$, both functions are equal to unity. These functions are shown in Fig.~\ref{fig:heraldedG2} for the case of a multimode PDC. 

We see in Fig.~\ref{fig:heraldedG2} that, on the one hand, the single-heralded autocorrelation function has the same shape as the non-heralded one shown in Fig.~\ref{fig:KTP}(d), corresponding to photon bunching. On the other hand, the double-heralded autocorrelation function has values much lower than unity for delays not exceeding the pump period, which is a signature of antibunching. However, on the timescale of the coherence time, this function exhibits a local maximum, corresponding to photon bunching. One may say that the double-heralded function is a witness of light nonclassicality, while the single-heralded function is more appropriate for the description of photon bunching. The state of the ordinary wave upon the heralding event is nonclassical because the vacuum is removed \cite{Lee95,Horoshko19tds}, but the removal of the vacuum does not affect the bunching property of the photons when two of them are present in the field.

The difference between the two autocorrelation functions can be further understood by rewriting Eq. (\ref{dheralded}) as $g_\text{dh}^{(2)}(\tau) = g_\text{sh}^{(2)}(\tau)/g_\text{int}^{(1,1)}(\tau)$, where 
\begin{equation}
g_\text{int}^{(1,1)}(\tau) = \frac{P_{2|h}}{P_2} = \frac
    {\left\langle :\hat I_o(t+\tau)\int_{\Pi_t} \hat I_e(t')dt':\right\rangle}
    {\langle \hat I_o(t+\tau)\rangle \left\langle \int_{\Pi_t} \hat I_e(t')dt'\right\rangle}    
\end{equation}
is the integrated cross-correlation function showing how many times the conditional probability of detection in interval 2 (under the condition of a heralding event in period $\Pi_t$) exceeds the unconditional one. From Eqs.~(\ref{BogoliubovA}) and (\ref{BogoliubovB}) we obtain
\begin{equation}
g_\text{int}^{(1,1)}(\tau) =
    \begin{cases}
        1+\frac1{P_b},& \text{for } t+\tau\in \Pi_t,\\ \\
        1,& \text{for}\ t+\tau\notin \Pi_t,
    \end{cases}
\end{equation}
which shows that within the same period as the heralding event, the conditional probability is $1+P_b^{-1}$ times higher than the unconditional one.

\section{Conclusion \label{sec:Conclusion}}

We have shown how the measure of entanglement of two photons generated in a nondegenerate PDC, the Schmidt number, is related to the degree of global coherence of one of these photons. This relation provides a method for determining the number of modes of photon pairs in two entangled beams by measuring the time-magnified $g^{(2)}(\tau)$ function of one beam. The proposed method is most efficient in the multimode case and, in this sense, is complementary to the traditional measurement of the time-integrated second-order autocorrelation function $g^{(2)}_\text{int}$, which enables the determination of the number of modes when it is relatively low \cite{Christ11}. Extending the consideration to the generation of heralded photons, we predicted that the time-magnified double-heralded autocorrelation function exhibits a local maximum around zero delay, corresponding to photon bunching. The width of this function can also be used to find the degree of global coherence. The new method can be applied to the characterization of multimode sources of entangled photons used for high-dimensional time-frequency encoding in photonic quantum technologies.

\section*{Acknowledgments} 
This work was supported by the QuantERA network (projects QuICHE and EXTRASENS), which has received funding from the European Union's Horizon 2020 research and innovation programme under grant agreement no. 731473. It was funded by Agence Nationale de la Recherche (France), grant ANR-19-QUANT-0001, the Federal Ministry of Education and Research of Germany, grant 13N16935, and the National Science Centre (Poland), projects no. 2019/32/Z/ST2/00018 and no. 2023/50/E/ST2/00703. M.M. and M.K. acknowledge the support from the University of Warsaw within the ``Excellence Initiative -- Research University'' framework.

\appendix
\section{Gaussian model \label{sec:appendixa}}
Substituting Eq. \eqref{JSAgauss} into \eqref{JtildeDef}, we arrive at
\begin{align}\label{Jsubst}
    \Tilde{J}_0 (t,t') &= \int\int \gamma L \alpha_0 \exp\left[-\frac{(\Omega+\Omega')^2}{4\Omega_p^2}-\frac{(\tau_o\Omega+\tau_e\Omega')^2}{2\sigma_s^2}\right]\nonumber\\
    &\quad\times e^{-i\Omega t - i\Omega't'} \frac{\mathrm{d}\Omega\mathrm{d}\Omega'}{(2\pi)^2}.
\end{align}
This integration can be taken analytically by the multidimensional Gaussian integration formula
\begin{equation}\label{multiGauss}
    \int_{-\infty}^{\infty} e^{-\frac12\mathbf{u}^T\boldsymbol{\Lambda} \mathbf{u}+i\mathbf{v}^T\mathbf{u}} d^n u=\left[\frac{(2\pi)^n}{\det\Lambda}\right]^{\frac12} e^{-\frac12\mathbf{v}^T\boldsymbol{\Lambda}^{-1} \mathbf{v}},
\end{equation}
where $\mathbf{u}$ and $\mathbf{v}$ are two $n$-dimensional column vectors, while $\boldsymbol{\Lambda}$ is a symmetric $n\times n$ matrix. In this way, Eq. \eqref{Jsubst} obtains the form
\begin{equation}
    \Tilde{J}_0 (t,t') =  \frac{ \gamma L \alpha_0}{2\pi\sqrt{\det\boldsymbol{\Lambda}}}  e^{-\frac12\mathbf{v}^T\boldsymbol{\Lambda}^{-1} \mathbf{v}},
\end{equation}
where
\begin{equation}\label{Lambda}
\boldsymbol{\Lambda}=\frac{1}{2\Omega_{p}^{2}}\left(
\begin{array}{cc}
 1+T^{2}_{o} & 1+T_o T_e  \\
 1+T_o T_e & 1+T^{2}_{e}  \\
\end{array}
\right),
\end{equation}
and $\textbf{v}=(-t,-t')^T$. From this, we find $\textbf{M}=\boldsymbol{\Lambda}^{-1}/2$ as specified by Eq. (\ref{M}) and obtain Eq. \eqref{JtildeGauss}.

Now, substituting Eq. \eqref{JtildeGauss} into Eq. \eqref{Gddef}, we arrive at
\begin{eqnarray}\label{GJTA}
&&G^{(1)}_{d}(t,\tau) = \mathcal{E}_o^2 J_1^2\\\nonumber
&&\times\int e^{- M_{11} t^2 - 2 M_{22} t'^2 - 2 M_{12}tt'- M_{11} (t+\tau)^2 - 2 M_{12}(t+\tau)t'} \mathrm{d}t'\\\nonumber
&&= \frac{\sqrt{\pi}\mathcal{E}_o^2 J_1^2 }{\sqrt{2M_{22}}} e^{- M_{11} t^2 - M_{11} (t+\tau)^2 + M^{2}_{12}(2t+\tau)^2/2M_{22}},
\end{eqnarray}
where we have used the one-dimensional version of Eq. (\ref{multiGauss}).

Putting $\tau=0$ in Eq. (\ref{GJTA}), we obtain the mean intensity of the ordinary wave
\begin{equation}\label{Ioapp}
I_d(t) = \frac{\sqrt{\pi}\mathcal{E}_o^2 J_1^2 }{\sqrt{2M_{22}}} e^{- 2M_{11} t^2 + 2M^{2}_{12}t^2/M_{22}},
\end{equation}
which we rewrite in the form of Eq. (\ref{Imu}) with  $\Delta t_o^2=M_{22}\det\boldsymbol{\Lambda}$ and $P_b = \pi J_1^2\Delta t_o/\sqrt{M_{22}}$. Substituting the matrix elements from Eqs. (\ref{Lambda}) and (\ref{M}), we obtain Eqs. (\ref{dt}) and (\ref{Pb}).

Comparing Eqs. \eqref{GJTA} and \eqref{Ioapp}, we arrive at Eq. \eqref{Gd} with $\Delta\tau_o^2=M_{22}/M_{12}^2$, which is equivalent to Eq. (\ref{dtau}).
\\

\section{Temporal Schmidt decomposition for a type-II biphoton \label{sec:appendixb}}

The spectral decomposition of a real symmetric double-Gaussian kernel is given by multiplying both sides of the Mehler's formula for Hermite polynomials by $e^{-x^2/2-y^2/2}$ \cite{Grice01,Horoshko19}
\begin{equation}\label{Mehler}
\frac1{\sqrt{\pi}}e^{-\frac{1+q^2}{2(1-q^2)}\left(x^2+y^2\right)+\frac{2q}{1-q^2}xy}
= \sum_{n=0}^\infty pq^n h_n(x) h_n(y),
\end{equation}
where $-1<q<1$, $p=\sqrt{1-q^2}$, and $h_n(x) = \left(2^nn!\sqrt{\pi}\right)^{-\frac12}H_n(x)e^{-x^2/2}$ is the Hermite-Gauss function, $H_n(x)$ being the Hermite polynomial. Identifying $x=t/\tau_1$, $y=t'\tau_2$, $q=\sqrt{(K-1)/(K+1)}$, and equalizing the exponents on the right-hand side of Eq. (\ref{JtildeGauss}) and the left-hand side of Eq. (\ref{Mehler}), we find the following system of equations:
\begin{eqnarray}
\frac{K}{\tau_1^2} &=& 2\Omega_p^2\frac{1+T_e^2}{(T_o-T_e)^2},\\
\frac{K}{\tau_2^2} &=& 2\Omega_p^2\frac{1+T_o^2}{(T_o-T_e)^2},\\
\frac{\sqrt{K^2-1}}{\tau_1\tau_2} &=& 2\Omega_p^2\frac{1+T_oT_e}{(T_o-T_e)^2}.
\end{eqnarray}
Solving this system for $K$, $\tau_1$ and $\tau_2$, we obtain Eqs. (\ref{K}) and (\ref{tau12}). We imply here that $1+T_oT_e>0$, which is the regime of our interest.

The singular value (Schmidt) decomposition of JTA can be easily obtained from its spectral decomposition, especially in the case of a positive $q$, considered here. Since all eigenvalues $pq^n$ are positive, they are also the singular values. The corresponding singular functions are given by the eigenfunctions, Eqs. (\ref{psin}) and (\ref{phin}). The factor before the sum in Eq. (\ref{Schmidt}) is given by $\sqrt{\pi\tau_1\tau_2}J_1$, which coincides with $\sqrt{P_b}$.

\bibliography{Temporal-Imaging2023}

\begin{thebibliography}{66}%
\makeatletter
\providecommand \@ifxundefined [1]{%
 \@ifx{#1\undefined}
}%
\providecommand \@ifnum [1]{%
 \ifnum #1\expandafter \@firstoftwo
 \else \expandafter \@secondoftwo
 \fi
}%
\providecommand \@ifx [1]{%
 \ifx #1\expandafter \@firstoftwo
 \else \expandafter \@secondoftwo
 \fi
}%
\providecommand \natexlab [1]{#1}%
\providecommand \enquote  [1]{``#1''}%
\providecommand \bibnamefont  [1]{#1}%
\providecommand \bibfnamefont [1]{#1}%
\providecommand \citenamefont [1]{#1}%
\providecommand \href@noop [0]{\@secondoftwo}%
\providecommand \href [0]{\begingroup \@sanitize@url \@href}%
\providecommand \@href[1]{\@@startlink{#1}\@@href}%
\providecommand \@@href[1]{\endgroup#1\@@endlink}%
\providecommand \@sanitize@url [0]{\catcode `\\12\catcode `\$12\catcode
  `\&12\catcode `\#12\catcode `\^12\catcode `\_12\catcode `\%12\relax}%
\providecommand \@@startlink[1]{}%
\providecommand \@@endlink[0]{}%
\providecommand \url  [0]{\begingroup\@sanitize@url \@url }%
\providecommand \@url [1]{\endgroup\@href {#1}{\urlprefix }}%
\providecommand \urlprefix  [0]{URL }%
\providecommand \Eprint [0]{\href }%
\providecommand \doibase [0]{https://doi.org/}%
\providecommand \selectlanguage [0]{\@gobble}%
\providecommand \bibinfo  [0]{\@secondoftwo}%
\providecommand \bibfield  [0]{\@secondoftwo}%
\providecommand \translation [1]{[#1]}%
\providecommand \BibitemOpen [0]{}%
\providecommand \bibitemStop [0]{}%
\providecommand \bibitemNoStop [0]{.\EOS\space}%
\providecommand \EOS [0]{\spacefactor3000\relax}%
\providecommand \BibitemShut  [1]{\csname bibitem#1\endcsname}%
\let\auto@bib@innerbib\@empty
\bibitem [{\citenamefont {Glauber}(1963)}]{Glauber63}%
  \BibitemOpen
  \bibfield  {author} {\bibinfo {author} {\bibfnamefont {R.~J.}\ \bibnamefont
  {Glauber}},\ }\bibfield  {title} {\bibinfo {title} {The quantum theory of
  optical coherence},\ }\href {https://doi.org/10.1103/PhysRev.130.2529}
  {\bibfield  {journal} {\bibinfo  {journal} {Phys. Rev.}\ }\textbf {\bibinfo
  {volume} {130}},\ \bibinfo {pages} {2529} (\bibinfo {year}
  {1963})}\BibitemShut {NoStop}%
\bibitem [{\citenamefont {Mandel}\ and\ \citenamefont
  {Wolf}(1995)}]{Mandel&Wolf}%
  \BibitemOpen
  \bibfield  {author} {\bibinfo {author} {\bibfnamefont {L.}~\bibnamefont
  {Mandel}}\ and\ \bibinfo {author} {\bibfnamefont {E.}~\bibnamefont {Wolf}},\
  }\href {https://doi.org/10.1017/CBO9781139644105} {\emph {\bibinfo {title}
  {Optical Coherence and Quantum Optics}}}\ (\bibinfo  {publisher} {Cambridge
  University Press},\ \bibinfo {address} {Cambridge},\ \bibinfo {year}
  {1995})\BibitemShut {NoStop}%
\bibitem [{\citenamefont {Arakawa}\ and\ \citenamefont
  {Holmes}(2020)}]{Arakawa20}%
  \BibitemOpen
  \bibfield  {author} {\bibinfo {author} {\bibfnamefont {Y.}~\bibnamefont
  {Arakawa}}\ and\ \bibinfo {author} {\bibfnamefont {M.~J.}\ \bibnamefont
  {Holmes}},\ }\bibfield  {title} {\bibinfo {title} {{Progress in quantum-dot
  single photon sources for quantum information technologies: A broad spectrum
  overview}},\ }\href {https://doi.org/10.1063/5.0010193} {\bibfield  {journal}
  {\bibinfo  {journal} {Appl. Phys. Rev.}\ }\textbf {\bibinfo {volume} {7}},\
  \bibinfo {pages} {021309} (\bibinfo {year} {2020})}\BibitemShut {NoStop}%
\bibitem [{\citenamefont {Toninelli}\ \emph {et~al.}(2021)\citenamefont
  {Toninelli}, \citenamefont {Gerhardt}, \citenamefont {Clark}, \citenamefont
  {Reserbat-Plantey}, \citenamefont {G{\"o}tzinger}, \citenamefont
  {Ristanovi{\'c}}, \citenamefont {Colautti}, \citenamefont {Lombardi},
  \citenamefont {Major}, \citenamefont {Deperasi{\'n}ska} \emph
  {et~al.}}]{Toninelli21}%
  \BibitemOpen
  \bibfield  {author} {\bibinfo {author} {\bibfnamefont {C.}~\bibnamefont
  {Toninelli}}, \bibinfo {author} {\bibfnamefont {I.}~\bibnamefont {Gerhardt}},
  \bibinfo {author} {\bibfnamefont {A.}~\bibnamefont {Clark}}, \bibinfo
  {author} {\bibfnamefont {A.}~\bibnamefont {Reserbat-Plantey}}, \bibinfo
  {author} {\bibfnamefont {S.}~\bibnamefont {G{\"o}tzinger}}, \bibinfo {author}
  {\bibfnamefont {Z.}~\bibnamefont {Ristanovi{\'c}}}, \bibinfo {author}
  {\bibfnamefont {M.}~\bibnamefont {Colautti}}, \bibinfo {author}
  {\bibfnamefont {P.}~\bibnamefont {Lombardi}}, \bibinfo {author}
  {\bibfnamefont {K.}~\bibnamefont {Major}}, \bibinfo {author} {\bibfnamefont
  {I.}~\bibnamefont {Deperasi{\'n}ska}}, \emph {et~al.},\ }\bibfield  {title}
  {\bibinfo {title} {Single organic molecules for photonic quantum
  technologies},\ }\href {https://doi.org/10.1038/s41563-021-00987-4}
  {\bibfield  {journal} {\bibinfo  {journal} {Nat. Materials}\ }\textbf
  {\bibinfo {volume} {20}},\ \bibinfo {pages} {1615} (\bibinfo {year}
  {2021})}\BibitemShut {NoStop}%
\bibitem [{\citenamefont {Doherty}\ \emph {et~al.}(2013)\citenamefont
  {Doherty}, \citenamefont {Manson}, \citenamefont {Delaney}, \citenamefont
  {Jelezko}, \citenamefont {Wrachtrup},\ and\ \citenamefont
  {Hollenberg}}]{Doherty13}%
  \BibitemOpen
  \bibfield  {author} {\bibinfo {author} {\bibfnamefont {M.~W.}\ \bibnamefont
  {Doherty}}, \bibinfo {author} {\bibfnamefont {N.~B.}\ \bibnamefont {Manson}},
  \bibinfo {author} {\bibfnamefont {P.}~\bibnamefont {Delaney}}, \bibinfo
  {author} {\bibfnamefont {F.}~\bibnamefont {Jelezko}}, \bibinfo {author}
  {\bibfnamefont {J.}~\bibnamefont {Wrachtrup}},\ and\ \bibinfo {author}
  {\bibfnamefont {L.~C.}\ \bibnamefont {Hollenberg}},\ }\bibfield  {title}
  {\bibinfo {title} {The nitrogen-vacancy colour centre in diamond},\ }\href
  {https://doi.org/https://doi.org/10.1016/j.physrep.2013.02.001} {\bibfield
  {journal} {\bibinfo  {journal} {Phys. Rep.}\ }\textbf {\bibinfo {volume}
  {528}},\ \bibinfo {pages} {1 } (\bibinfo {year} {2013})}\BibitemShut
  {NoStop}%
\bibitem [{\citenamefont {Christ}\ \emph {et~al.}(2011)\citenamefont {Christ},
  \citenamefont {Laiho}, \citenamefont {Eckstein}, \citenamefont {Cassemiro},\
  and\ \citenamefont {Silberhorn}}]{Christ11}%
  \BibitemOpen
  \bibfield  {author} {\bibinfo {author} {\bibfnamefont {A.}~\bibnamefont
  {Christ}}, \bibinfo {author} {\bibfnamefont {K.}~\bibnamefont {Laiho}},
  \bibinfo {author} {\bibfnamefont {A.}~\bibnamefont {Eckstein}}, \bibinfo
  {author} {\bibfnamefont {K.~N.}\ \bibnamefont {Cassemiro}},\ and\ \bibinfo
  {author} {\bibfnamefont {C.}~\bibnamefont {Silberhorn}},\ }\bibfield  {title}
  {\bibinfo {title} {Probing multimode squeezing with correlation functions},\
  }\href {https://doi.org/10.1088/1367-2630/13/3/033027} {\bibfield  {journal}
  {\bibinfo  {journal} {New J. Phys.}\ }\textbf {\bibinfo {volume} {13}},\
  \bibinfo {pages} {033027} (\bibinfo {year} {2011})}\BibitemShut {NoStop}%
\bibitem [{\citenamefont {Kolner}(1994)}]{Kolner94}%
  \BibitemOpen
  \bibfield  {author} {\bibinfo {author} {\bibfnamefont {B.~H.}\ \bibnamefont
  {Kolner}},\ }\bibfield  {title} {\bibinfo {title} {Space-time duality and the
  theory of temporal imaging},\ }\href@noop {} {\bibfield  {journal} {\bibinfo
  {journal} {IEEE J. Quantum Elect.}\ }\textbf {\bibinfo {volume} {30}},\
  \bibinfo {pages} {1951} (\bibinfo {year} {1994})}\BibitemShut {NoStop}%
\bibitem [{\citenamefont {Salem}\ \emph {et~al.}(2013)\citenamefont {Salem},
  \citenamefont {Foster},\ and\ \citenamefont {Gaeta}}]{Salem13}%
  \BibitemOpen
  \bibfield  {author} {\bibinfo {author} {\bibfnamefont {R.}~\bibnamefont
  {Salem}}, \bibinfo {author} {\bibfnamefont {M.~A.}\ \bibnamefont {Foster}},\
  and\ \bibinfo {author} {\bibfnamefont {A.~L.}\ \bibnamefont {Gaeta}},\
  }\bibfield  {title} {\bibinfo {title} {Application of space–time duality to
  ultrahigh-speed optical signal processing},\ }\href@noop {} {\bibfield
  {journal} {\bibinfo  {journal} {Adv. Opt. Photonics}\ }\textbf {\bibinfo
  {volume} {5}},\ \bibinfo {pages} {275} (\bibinfo {year} {2013})}\BibitemShut
  {NoStop}%
\bibitem [{\citenamefont {Lavoie}\ \emph {et~al.}(2013)\citenamefont {Lavoie},
  \citenamefont {Donohue}, \citenamefont {Wright}, \citenamefont {Fedrizzi},\
  and\ \citenamefont {Resch}}]{Lavoie13}%
  \BibitemOpen
  \bibfield  {author} {\bibinfo {author} {\bibfnamefont {J.}~\bibnamefont
  {Lavoie}}, \bibinfo {author} {\bibfnamefont {J.~M.}\ \bibnamefont {Donohue}},
  \bibinfo {author} {\bibfnamefont {L.~G.}\ \bibnamefont {Wright}}, \bibinfo
  {author} {\bibfnamefont {A.}~\bibnamefont {Fedrizzi}},\ and\ \bibinfo
  {author} {\bibfnamefont {K.~J.}\ \bibnamefont {Resch}},\ }\bibfield  {title}
  {\bibinfo {title} {Spectral compression of single photons},\ }\href@noop {}
  {\bibfield  {journal} {\bibinfo  {journal} {Nat. Photonics}\ }\textbf
  {\bibinfo {volume} {7}},\ \bibinfo {pages} {363} (\bibinfo {year}
  {2013})}\BibitemShut {NoStop}%
\bibitem [{\citenamefont {Karpiński}\ \emph {et~al.}(2017)\citenamefont
  {Karpiński}, \citenamefont {Jachura}, \citenamefont {Wright},\ and\
  \citenamefont {Smith}}]{Karpinski17}%
  \BibitemOpen
  \bibfield  {author} {\bibinfo {author} {\bibfnamefont {M.}~\bibnamefont
  {Karpiński}}, \bibinfo {author} {\bibfnamefont {M.}~\bibnamefont {Jachura}},
  \bibinfo {author} {\bibfnamefont {L.~J.}\ \bibnamefont {Wright}},\ and\
  \bibinfo {author} {\bibfnamefont {B.~J.}\ \bibnamefont {Smith}},\ }\bibfield
  {title} {\bibinfo {title} {Bandwidth manipulation of quantum light by an
  electro-optic time lens},\ }\href@noop {} {\bibfield  {journal} {\bibinfo
  {journal} {Nat. Photonics}\ }\textbf {\bibinfo {volume} {11}},\ \bibinfo
  {pages} {53–57} (\bibinfo {year} {2017})}\BibitemShut {NoStop}%
\bibitem [{\citenamefont {Mittal}\ \emph {et~al.}(2017)\citenamefont {Mittal},
  \citenamefont {Orre}, \citenamefont {Restelli}, \citenamefont {Salem},
  \citenamefont {Goldschmidt},\ and\ \citenamefont {Hafezi}}]{Mittal17}%
  \BibitemOpen
  \bibfield  {author} {\bibinfo {author} {\bibfnamefont {S.}~\bibnamefont
  {Mittal}}, \bibinfo {author} {\bibfnamefont {V.~V.}\ \bibnamefont {Orre}},
  \bibinfo {author} {\bibfnamefont {A.}~\bibnamefont {Restelli}}, \bibinfo
  {author} {\bibfnamefont {R.}~\bibnamefont {Salem}}, \bibinfo {author}
  {\bibfnamefont {E.~A.}\ \bibnamefont {Goldschmidt}},\ and\ \bibinfo {author}
  {\bibfnamefont {M.}~\bibnamefont {Hafezi}},\ }\bibfield  {title} {\bibinfo
  {title} {Temporal and spectral manipulations of correlated photons using a
  time lens},\ }\href {https://doi.org/10.1103/PhysRevA.96.043807} {\bibfield
  {journal} {\bibinfo  {journal} {Phys. Rev. A}\ }\textbf {\bibinfo {volume}
  {96}},\ \bibinfo {pages} {043807} (\bibinfo {year} {2017})}\BibitemShut
  {NoStop}%
\bibitem [{\citenamefont {Patera}\ \emph {et~al.}(2017)\citenamefont {Patera},
  \citenamefont {Shi}, \citenamefont {Horoshko},\ and\ \citenamefont
  {Kolobov}}]{Patera17}%
  \BibitemOpen
  \bibfield  {author} {\bibinfo {author} {\bibfnamefont {G.}~\bibnamefont
  {Patera}}, \bibinfo {author} {\bibfnamefont {J.}~\bibnamefont {Shi}},
  \bibinfo {author} {\bibfnamefont {D.~B.}\ \bibnamefont {Horoshko}},\ and\
  \bibinfo {author} {\bibfnamefont {M.~I.}\ \bibnamefont {Kolobov}},\
  }\bibfield  {title} {\bibinfo {title} {Quantum temporal imaging: application
  of a time lens to quantum optics},\ }\href@noop {} {\bibfield  {journal}
  {\bibinfo  {journal} {J. Opt.}\ }\textbf {\bibinfo {volume} {19}},\ \bibinfo
  {pages} {054001} (\bibinfo {year} {2017})}\BibitemShut {NoStop}%
\bibitem [{\citenamefont {Patera}\ \emph {et~al.}(2018)\citenamefont {Patera},
  \citenamefont {Horoshko},\ and\ \citenamefont {Kolobov}}]{Patera18}%
  \BibitemOpen
  \bibfield  {author} {\bibinfo {author} {\bibfnamefont {G.}~\bibnamefont
  {Patera}}, \bibinfo {author} {\bibfnamefont {D.~B.}\ \bibnamefont
  {Horoshko}},\ and\ \bibinfo {author} {\bibfnamefont {M.~I.}\ \bibnamefont
  {Kolobov}},\ }\bibfield  {title} {\bibinfo {title} {Space-time duality and
  quantum temporal imaging},\ }\href
  {https://doi.org/10.1103/PhysRevA.98.053815} {\bibfield  {journal} {\bibinfo
  {journal} {Phys. Rev. A}\ }\textbf {\bibinfo {volume} {98}},\ \bibinfo
  {pages} {053815} (\bibinfo {year} {2018})}\BibitemShut {NoStop}%
\bibitem [{\citenamefont {Shi}\ \emph {et~al.}(2020)\citenamefont {Shi},
  \citenamefont {Patera}, \citenamefont {Horoshko},\ and\ \citenamefont
  {Kolobov}}]{Shi20}%
  \BibitemOpen
  \bibfield  {author} {\bibinfo {author} {\bibfnamefont {J.}~\bibnamefont
  {Shi}}, \bibinfo {author} {\bibfnamefont {G.}~\bibnamefont {Patera}},
  \bibinfo {author} {\bibfnamefont {D.~B.}\ \bibnamefont {Horoshko}},\ and\
  \bibinfo {author} {\bibfnamefont {M.~I.}\ \bibnamefont {Kolobov}},\
  }\bibfield  {title} {\bibinfo {title} {Quantum temporal imaging of
  antibunching},\ }\href {https://doi.org/10.1364/JOSAB.400270} {\bibfield
  {journal} {\bibinfo  {journal} {J. Opt. Soc. Am. B}\ }\textbf {\bibinfo
  {volume} {37}},\ \bibinfo {pages} {3741} (\bibinfo {year}
  {2020})}\BibitemShut {NoStop}%
\bibitem [{\citenamefont {So{\'s}nicki}\ \emph {et~al.}(2020)\citenamefont
  {So{\'s}nicki}, \citenamefont {Miko{\l}ajczyk}, \citenamefont {Golestani},\
  and\ \citenamefont {Karpi{\'n}ski}}]{Sosnicki20}%
  \BibitemOpen
  \bibfield  {author} {\bibinfo {author} {\bibfnamefont {F.}~\bibnamefont
  {So{\'s}nicki}}, \bibinfo {author} {\bibfnamefont {M.}~\bibnamefont
  {Miko{\l}ajczyk}}, \bibinfo {author} {\bibfnamefont {A.}~\bibnamefont
  {Golestani}},\ and\ \bibinfo {author} {\bibfnamefont {M.}~\bibnamefont
  {Karpi{\'n}ski}},\ }\bibfield  {title} {\bibinfo {title} {{Aperiodic
  electro-optic time lens for spectral manipulation of single-photon pulses}},\
  }\href {https://doi.org/10.1063/5.0011077} {\bibfield  {journal} {\bibinfo
  {journal} {Appl. Phys. Lett.}\ }\textbf {\bibinfo {volume} {116}},\ \bibinfo
  {pages} {234003} (\bibinfo {year} {2020})}\BibitemShut {NoStop}%
\bibitem [{\citenamefont {Joshi}\ \emph {et~al.}(2022)\citenamefont {Joshi},
  \citenamefont {Sparkes}, \citenamefont {Farsi}, \citenamefont {Gerrits},
  \citenamefont {Verma}, \citenamefont {Ramelow}, \citenamefont {Nam},\ and\
  \citenamefont {Gaeta}}]{Joshi22}%
  \BibitemOpen
  \bibfield  {author} {\bibinfo {author} {\bibfnamefont {C.}~\bibnamefont
  {Joshi}}, \bibinfo {author} {\bibfnamefont {B.~M.}\ \bibnamefont {Sparkes}},
  \bibinfo {author} {\bibfnamefont {A.}~\bibnamefont {Farsi}}, \bibinfo
  {author} {\bibfnamefont {T.}~\bibnamefont {Gerrits}}, \bibinfo {author}
  {\bibfnamefont {V.}~\bibnamefont {Verma}}, \bibinfo {author} {\bibfnamefont
  {S.}~\bibnamefont {Ramelow}}, \bibinfo {author} {\bibfnamefont {S.~W.}\
  \bibnamefont {Nam}},\ and\ \bibinfo {author} {\bibfnamefont {A.~L.}\
  \bibnamefont {Gaeta}},\ }\bibfield  {title} {\bibinfo {title}
  {Picosecond-resolution single-photon time lens for temporal mode quantum
  processing},\ }\href {https://doi.org/10.1364/OPTICA.439827} {\bibfield
  {journal} {\bibinfo  {journal} {Optica}\ }\textbf {\bibinfo {volume} {9}},\
  \bibinfo {pages} {364} (\bibinfo {year} {2022})}\BibitemShut {NoStop}%
\bibitem [{\citenamefont {So{\'s}nicki}\ \emph {et~al.}(2023)\citenamefont
  {So{\'s}nicki}, \citenamefont {Miko{\l}ajczyk}, \citenamefont {Golestani},\
  and\ \citenamefont {Karpi{\'n}ski}}]{Sosnicki23}%
  \BibitemOpen
  \bibfield  {author} {\bibinfo {author} {\bibfnamefont {F.}~\bibnamefont
  {So{\'s}nicki}}, \bibinfo {author} {\bibfnamefont {M.}~\bibnamefont
  {Miko{\l}ajczyk}}, \bibinfo {author} {\bibfnamefont {A.}~\bibnamefont
  {Golestani}},\ and\ \bibinfo {author} {\bibfnamefont {M.}~\bibnamefont
  {Karpi{\'n}ski}},\ }\bibfield  {title} {\bibinfo {title} {Interface between
  picosecond and nanosecond quantum light pulses},\ }\href
  {https://doi.org/10.1038/s41566-023-01214-z} {\bibfield  {journal} {\bibinfo
  {journal} {Nat. Photonics}\ }\textbf {\bibinfo {volume} {17}},\ \bibinfo
  {pages} {761} (\bibinfo {year} {2023})}\BibitemShut {NoStop}%
\bibitem [{\citenamefont {Klyshko}(1988)}]{KlyshkoBook}%
  \BibitemOpen
  \bibfield  {author} {\bibinfo {author} {\bibfnamefont {D.~N.}\ \bibnamefont
  {Klyshko}},\ }\href@noop {} {\emph {\bibinfo {title} {Photons and Nonlinear
  Optics}}}\ (\bibinfo  {publisher} {Gordon and Breach},\ \bibinfo {address}
  {New York},\ \bibinfo {year} {1988})\BibitemShut {NoStop}%
\bibitem [{\citenamefont {Rubin}\ \emph {et~al.}(1994)\citenamefont {Rubin},
  \citenamefont {Klyshko}, \citenamefont {Shih},\ and\ \citenamefont
  {Sergienko}}]{Rubin94}%
  \BibitemOpen
  \bibfield  {author} {\bibinfo {author} {\bibfnamefont {M.~H.}\ \bibnamefont
  {Rubin}}, \bibinfo {author} {\bibfnamefont {D.~N.}\ \bibnamefont {Klyshko}},
  \bibinfo {author} {\bibfnamefont {Y.~H.}\ \bibnamefont {Shih}},\ and\
  \bibinfo {author} {\bibfnamefont {A.~V.}\ \bibnamefont {Sergienko}},\
  }\bibfield  {title} {\bibinfo {title} {Theory of two-photon entanglement in
  {type-II} optical parametric down-conversion},\ }\href
  {https://doi.org/10.1103/PhysRevA.50.5122} {\bibfield  {journal} {\bibinfo
  {journal} {Phys. Rev. A}\ }\textbf {\bibinfo {volume} {50}},\ \bibinfo
  {pages} {5122} (\bibinfo {year} {1994})}\BibitemShut {NoStop}%
\bibitem [{\citenamefont {Grice}\ and\ \citenamefont
  {Walmsley}(1997)}]{Grice97}%
  \BibitemOpen
  \bibfield  {author} {\bibinfo {author} {\bibfnamefont {W.~P.}\ \bibnamefont
  {Grice}}\ and\ \bibinfo {author} {\bibfnamefont {I.~A.}\ \bibnamefont
  {Walmsley}},\ }\bibfield  {title} {\bibinfo {title} {Spectral information and
  distinguishability in {type-II} down-conversion with a broadband pump},\
  }\href {https://doi.org/10.1103/PhysRevA.56.1627} {\bibfield  {journal}
  {\bibinfo  {journal} {Phys. Rev. A}\ }\textbf {\bibinfo {volume} {56}},\
  \bibinfo {pages} {1627} (\bibinfo {year} {1997})}\BibitemShut {NoStop}%
\bibitem [{\citenamefont {Keller}\ and\ \citenamefont
  {Rubin}(1997)}]{Keller97}%
  \BibitemOpen
  \bibfield  {author} {\bibinfo {author} {\bibfnamefont {T.~E.}\ \bibnamefont
  {Keller}}\ and\ \bibinfo {author} {\bibfnamefont {M.~H.}\ \bibnamefont
  {Rubin}},\ }\bibfield  {title} {\bibinfo {title} {Theory of two-photon
  entanglement for spontaneous parametric down-conversion driven by a narrow
  pump pulse},\ }\href {https://doi.org/10.1103/PhysRevA.56.1534} {\bibfield
  {journal} {\bibinfo  {journal} {Phys. Rev. A}\ }\textbf {\bibinfo {volume}
  {56}},\ \bibinfo {pages} {1534} (\bibinfo {year} {1997})}\BibitemShut
  {NoStop}%
\bibitem [{\citenamefont {Kolobov}(1999)}]{Kolobov99}%
  \BibitemOpen
  \bibfield  {author} {\bibinfo {author} {\bibfnamefont {M.~I.}\ \bibnamefont
  {Kolobov}},\ }\bibfield  {title} {\bibinfo {title} {The spatial behavior of
  nonclassical light},\ }\href@noop {} {\bibfield  {journal} {\bibinfo
  {journal} {Rev. Mod. Phys.}\ }\textbf {\bibinfo {volume} {71}},\ \bibinfo
  {pages} {1539} (\bibinfo {year} {1999})}\BibitemShut {NoStop}%
\bibitem [{\citenamefont {Huttner}\ \emph {et~al.}(1990)\citenamefont
  {Huttner}, \citenamefont {Serulnik},\ and\ \citenamefont
  {Ben-Aryeh}}]{Huttner90}%
  \BibitemOpen
  \bibfield  {author} {\bibinfo {author} {\bibfnamefont {B.}~\bibnamefont
  {Huttner}}, \bibinfo {author} {\bibfnamefont {S.}~\bibnamefont {Serulnik}},\
  and\ \bibinfo {author} {\bibfnamefont {Y.}~\bibnamefont {Ben-Aryeh}},\
  }\bibfield  {title} {\bibinfo {title} {Quantum analysis of light propagation
  in a parametric amplifier},\ }\href
  {https://doi.org/10.1103/PhysRevA.42.5594} {\bibfield  {journal} {\bibinfo
  {journal} {Phys. Rev. A}\ }\textbf {\bibinfo {volume} {42}},\ \bibinfo
  {pages} {5594} (\bibinfo {year} {1990})}\BibitemShut {NoStop}%
\bibitem [{\citenamefont {Horoshko}(2022)}]{Horoshko22}%
  \BibitemOpen
  \bibfield  {author} {\bibinfo {author} {\bibfnamefont {D.~B.}\ \bibnamefont
  {Horoshko}},\ }\bibfield  {title} {\bibinfo {title} {Generator of spatial
  evolution of the electromagnetic field},\ }\href
  {https://doi.org/10.1103/PhysRevA.105.013708} {\bibfield  {journal} {\bibinfo
   {journal} {Phys. Rev. A}\ }\textbf {\bibinfo {volume} {105}},\ \bibinfo
  {pages} {013708} (\bibinfo {year} {2022})}\BibitemShut {NoStop}%
\bibitem [{\citenamefont {Shen}(1967)}]{Shen67}%
  \BibitemOpen
  \bibfield  {author} {\bibinfo {author} {\bibfnamefont {Y.~R.}\ \bibnamefont
  {Shen}},\ }\bibfield  {title} {\bibinfo {title} {Quantum statistics of
  nonlinear optics},\ }\href {https://doi.org/10.1103/PhysRev.155.921}
  {\bibfield  {journal} {\bibinfo  {journal} {Phys. Rev.}\ }\textbf {\bibinfo
  {volume} {155}},\ \bibinfo {pages} {921} (\bibinfo {year}
  {1967})}\BibitemShut {NoStop}%
\bibitem [{\citenamefont {Horoshko}\ \emph {et~al.}(2024)\citenamefont
  {Horoshko}, \citenamefont {Kolobov}, \citenamefont {Parigi},\ and\
  \citenamefont {Treps}}]{Horoshko24}%
  \BibitemOpen
  \bibfield  {author} {\bibinfo {author} {\bibfnamefont {D.~B.}\ \bibnamefont
  {Horoshko}}, \bibinfo {author} {\bibfnamefont {M.~I.}\ \bibnamefont
  {Kolobov}}, \bibinfo {author} {\bibfnamefont {V.}~\bibnamefont {Parigi}},\
  and\ \bibinfo {author} {\bibfnamefont {N.}~\bibnamefont {Treps}},\ }\bibfield
   {title} {\bibinfo {title} {Few-mode squeezing in type-{I} parametric
  downconversion by complete group velocity matching},\ }\bibfield  {journal}
  {\bibinfo  {journal} {Opt. Lett.}\ }\href {https://doi.org/10.1364/OL.528280}
  {10.1364/OL.528280} (\bibinfo {year} {2024})\BibitemShut {NoStop}%
\bibitem [{\citenamefont {Boyd}(2020)}]{BoydBook}%
  \BibitemOpen
  \bibfield  {author} {\bibinfo {author} {\bibfnamefont {R.~W.}\ \bibnamefont
  {Boyd}},\ }\href@noop {} {\emph {\bibinfo {title} {Nonlinear Optics}}}\
  (\bibinfo  {publisher} {Academic Press},\ \bibinfo {address} {New York},\
  \bibinfo {year} {2020})\BibitemShut {NoStop}%
\bibitem [{\citenamefont {Horoshko}\ and\ \citenamefont
  {Kolobov}(2017)}]{Horoshko17}%
  \BibitemOpen
  \bibfield  {author} {\bibinfo {author} {\bibfnamefont {D.~B.}\ \bibnamefont
  {Horoshko}}\ and\ \bibinfo {author} {\bibfnamefont {M.~I.}\ \bibnamefont
  {Kolobov}},\ }\bibfield  {title} {\bibinfo {title} {Generation of monocycle
  squeezed light in chirped quasi-phase-matched nonlinear crystals},\ }\href
  {https://doi.org/10.1103/PhysRevA.95.033837} {\bibfield  {journal} {\bibinfo
  {journal} {Phys. Rev. A}\ }\textbf {\bibinfo {volume} {95}},\ \bibinfo
  {pages} {033837} (\bibinfo {year} {2017})}\BibitemShut {NoStop}%
\bibitem [{\citenamefont {Caves}\ and\ \citenamefont {Crouch}(1987)}]{Caves87}%
  \BibitemOpen
  \bibfield  {author} {\bibinfo {author} {\bibfnamefont {C.~M.}\ \bibnamefont
  {Caves}}\ and\ \bibinfo {author} {\bibfnamefont {D.~D.}\ \bibnamefont
  {Crouch}},\ }\bibfield  {title} {\bibinfo {title} {Quantum wideband
  traveling-wave analysis of a degenerate parametric amplifier},\ }\href@noop
  {} {\bibfield  {journal} {\bibinfo  {journal} {J. Opt. Soc. Am. B}\ }\textbf
  {\bibinfo {volume} {4}},\ \bibinfo {pages} {1535} (\bibinfo {year}
  {1987})}\BibitemShut {NoStop}%
\bibitem [{\citenamefont {Telegin}\ and\ \citenamefont
  {Chirkin}(1985)}]{Telegin85}%
  \BibitemOpen
  \bibfield  {author} {\bibinfo {author} {\bibfnamefont {L.~S.}\ \bibnamefont
  {Telegin}}\ and\ \bibinfo {author} {\bibfnamefont {A.~S.}\ \bibnamefont
  {Chirkin}},\ }\bibfield  {title} {\bibinfo {title} {Reversal and
  reconstruction of the profile of ultrashort light pulses},\ }\href
  {https://doi.org/10.1070/QE1985v015n01ABEH005871} {\bibfield  {journal}
  {\bibinfo  {journal} {Sov. J. Quantum Electr.}\ }\textbf {\bibinfo {volume}
  {15}},\ \bibinfo {pages} {101} (\bibinfo {year} {1985})}\BibitemShut
  {NoStop}%
\bibitem [{\citenamefont {Kolner}\ and\ \citenamefont
  {Nazarathy}(1989)}]{Kolner89}%
  \BibitemOpen
  \bibfield  {author} {\bibinfo {author} {\bibfnamefont {B.~H.}\ \bibnamefont
  {Kolner}}\ and\ \bibinfo {author} {\bibfnamefont {M.}~\bibnamefont
  {Nazarathy}},\ }\bibfield  {title} {\bibinfo {title} {Temporal imaging with a
  time lens},\ }\href@noop {} {\bibfield  {journal} {\bibinfo  {journal} {Opt.
  Lett.}\ }\textbf {\bibinfo {volume} {14}},\ \bibinfo {pages} {630} (\bibinfo
  {year} {1989})}\BibitemShut {NoStop}%
\bibitem [{\citenamefont {Mouradian}\ \emph {et~al.}(2000)\citenamefont
  {Mouradian}, \citenamefont {Louradour}, \citenamefont {Messager},
  \citenamefont {Barthelemy},\ and\ \citenamefont {Froehly}}]{Mouradian00}%
  \BibitemOpen
  \bibfield  {author} {\bibinfo {author} {\bibfnamefont {L.}~\bibnamefont
  {Mouradian}}, \bibinfo {author} {\bibfnamefont {F.}~\bibnamefont
  {Louradour}}, \bibinfo {author} {\bibfnamefont {V.}~\bibnamefont {Messager}},
  \bibinfo {author} {\bibfnamefont {A.}~\bibnamefont {Barthelemy}},\ and\
  \bibinfo {author} {\bibfnamefont {C.}~\bibnamefont {Froehly}},\ }\bibfield
  {title} {\bibinfo {title} {Spectro-temporal imaging of femtosecond events},\
  }\href {https://doi.org/10.1109/3.848351} {\bibfield  {journal} {\bibinfo
  {journal} {IEEE Journal of Quantum Electronics}\ }\textbf {\bibinfo {volume}
  {36}},\ \bibinfo {pages} {795} (\bibinfo {year} {2000})}\BibitemShut
  {NoStop}%
\bibitem [{\citenamefont {Bennett}\ and\ \citenamefont
  {Kolner}(2000{\natexlab{a}})}]{Bennett00a}%
  \BibitemOpen
  \bibfield  {author} {\bibinfo {author} {\bibfnamefont {C.~V.}\ \bibnamefont
  {Bennett}}\ and\ \bibinfo {author} {\bibfnamefont {B.~H.}\ \bibnamefont
  {Kolner}},\ }\bibfield  {title} {\bibinfo {title} {Principles of parametric
  temporal imaging. {I}. {S}ystem configurations},\ }\href
  {https://doi.org/10.1109/3.831018} {\bibfield  {journal} {\bibinfo  {journal}
  {IEEE J. Quantum Elect.}\ }\textbf {\bibinfo {volume} {36}},\ \bibinfo
  {pages} {430} (\bibinfo {year} {2000}{\natexlab{a}})}\BibitemShut {NoStop}%
\bibitem [{\citenamefont {Bennett}\ and\ \citenamefont
  {Kolner}(2000{\natexlab{b}})}]{Bennett00b}%
  \BibitemOpen
  \bibfield  {author} {\bibinfo {author} {\bibfnamefont {C.~V.}\ \bibnamefont
  {Bennett}}\ and\ \bibinfo {author} {\bibfnamefont {B.~H.}\ \bibnamefont
  {Kolner}},\ }\bibfield  {title} {\bibinfo {title} {Principles of parametric
  temporal imaging. {II}. {S}ystem performance},\ }\href
  {https://doi.org/10.1109/3.845718} {\bibfield  {journal} {\bibinfo  {journal}
  {IEEE J. Quantum Elect.}\ }\textbf {\bibinfo {volume} {36}},\ \bibinfo
  {pages} {649} (\bibinfo {year} {2000}{\natexlab{b}})}\BibitemShut {NoStop}%
\bibitem [{\citenamefont {Foster}\ \emph {et~al.}(2008)\citenamefont {Foster},
  \citenamefont {Salem}, \citenamefont {Geraghty}, \citenamefont
  {Turner-Foster}, \citenamefont {Lipson},\ and\ \citenamefont
  {Gaeta}}]{Foster08}%
  \BibitemOpen
  \bibfield  {author} {\bibinfo {author} {\bibfnamefont {M.~A.}\ \bibnamefont
  {Foster}}, \bibinfo {author} {\bibfnamefont {R.}~\bibnamefont {Salem}},
  \bibinfo {author} {\bibfnamefont {D.~F.}\ \bibnamefont {Geraghty}}, \bibinfo
  {author} {\bibfnamefont {A.~C.}\ \bibnamefont {Turner-Foster}}, \bibinfo
  {author} {\bibfnamefont {M.}~\bibnamefont {Lipson}},\ and\ \bibinfo {author}
  {\bibfnamefont {A.~L.}\ \bibnamefont {Gaeta}},\ }\bibfield  {title} {\bibinfo
  {title} {Silicon-chip-based ultrafast optical oscilloscope},\ }\href@noop {}
  {\bibfield  {journal} {\bibinfo  {journal} {Nature}\ }\textbf {\bibinfo
  {volume} {456}},\ \bibinfo {pages} {81} (\bibinfo {year} {2008})}\BibitemShut
  {NoStop}%
\bibitem [{\citenamefont {Meir}\ \emph {et~al.}(2023)\citenamefont {Meir},
  \citenamefont {Tamir}, \citenamefont {Duadi}, \citenamefont {Cohen},\ and\
  \citenamefont {Fridman}}]{Meir23}%
  \BibitemOpen
  \bibfield  {author} {\bibinfo {author} {\bibfnamefont {S.}~\bibnamefont
  {Meir}}, \bibinfo {author} {\bibfnamefont {Y.}~\bibnamefont {Tamir}},
  \bibinfo {author} {\bibfnamefont {H.}~\bibnamefont {Duadi}}, \bibinfo
  {author} {\bibfnamefont {E.}~\bibnamefont {Cohen}},\ and\ \bibinfo {author}
  {\bibfnamefont {M.}~\bibnamefont {Fridman}},\ }\bibfield  {title} {\bibinfo
  {title} {Ultrafast temporal {SU(1,1)} interferometer},\ }\href
  {https://doi.org/10.1103/PhysRevLett.130.253601} {\bibfield  {journal}
  {\bibinfo  {journal} {Phys. Rev. Lett.}\ }\textbf {\bibinfo {volume} {130}},\
  \bibinfo {pages} {253601} (\bibinfo {year} {2023})}\BibitemShut {NoStop}%
\bibitem [{\citenamefont {Mazelanik}\ \emph {et~al.}(2020)\citenamefont
  {Mazelanik}, \citenamefont {Leszczy\'{n}ski}, \citenamefont {Lipka},
  \citenamefont {Parniak},\ and\ \citenamefont {Wasilewski}}]{Mazelanik20}%
  \BibitemOpen
  \bibfield  {author} {\bibinfo {author} {\bibfnamefont {M.}~\bibnamefont
  {Mazelanik}}, \bibinfo {author} {\bibfnamefont {A.}~\bibnamefont
  {Leszczy\'{n}ski}}, \bibinfo {author} {\bibfnamefont {M.}~\bibnamefont
  {Lipka}}, \bibinfo {author} {\bibfnamefont {M.}~\bibnamefont {Parniak}},\
  and\ \bibinfo {author} {\bibfnamefont {W.}~\bibnamefont {Wasilewski}},\
  }\bibfield  {title} {\bibinfo {title} {Temporal imaging for ultra-narrowband
  few-photon states of light},\ }\href {https://doi.org/10.1364/OPTICA.382891}
  {\bibfield  {journal} {\bibinfo  {journal} {Optica}\ }\textbf {\bibinfo
  {volume} {7}},\ \bibinfo {pages} {203} (\bibinfo {year} {2020})}\BibitemShut
  {NoStop}%
\bibitem [{\citenamefont {Mazelanik}\ \emph {et~al.}(2022)\citenamefont
  {Mazelanik}, \citenamefont {Leszczy{\'n}ski},\ and\ \citenamefont
  {Parniak}}]{Mazelanik22}%
  \BibitemOpen
  \bibfield  {author} {\bibinfo {author} {\bibfnamefont {M.}~\bibnamefont
  {Mazelanik}}, \bibinfo {author} {\bibfnamefont {A.}~\bibnamefont
  {Leszczy{\'n}ski}},\ and\ \bibinfo {author} {\bibfnamefont {M.}~\bibnamefont
  {Parniak}},\ }\bibfield  {title} {\bibinfo {title} {Optical-domain spectral
  super-resolution via a quantum-memory-based time-frequency processor},\
  }\href@noop {} {\bibfield  {journal} {\bibinfo  {journal} {Nature Comm.}\
  }\textbf {\bibinfo {volume} {13}},\ \bibinfo {pages} {1} (\bibinfo {year}
  {2022})}\BibitemShut {NoStop}%
\bibitem [{\citenamefont {Niewelt}\ \emph {et~al.}(2023)\citenamefont
  {Niewelt}, \citenamefont {Jastrz{e}bski}, \citenamefont {Kurzyna},
  \citenamefont {Nowosielski}, \citenamefont {Wasilewski}, \citenamefont
  {Mazelanik},\ and\ \citenamefont {Parniak}}]{Niewelt23}%
  \BibitemOpen
  \bibfield  {author} {\bibinfo {author} {\bibfnamefont {B.}~\bibnamefont
  {Niewelt}}, \bibinfo {author} {\bibfnamefont {M.}~\bibnamefont
  {Jastrz{e}bski}}, \bibinfo {author} {\bibfnamefont {S.}~\bibnamefont
  {Kurzyna}}, \bibinfo {author} {\bibfnamefont {J.}~\bibnamefont
  {Nowosielski}}, \bibinfo {author} {\bibfnamefont {W.}~\bibnamefont
  {Wasilewski}}, \bibinfo {author} {\bibfnamefont {M.}~\bibnamefont
  {Mazelanik}},\ and\ \bibinfo {author} {\bibfnamefont {M.}~\bibnamefont
  {Parniak}},\ }\bibfield  {title} {\bibinfo {title} {Experimental
  implementation of the optical fractional {Fourier} transform in the
  time-frequency domain},\ }\href
  {https://doi.org/10.1103/PhysRevLett.130.240801} {\bibfield  {journal}
  {\bibinfo  {journal} {Phys. Rev. Lett.}\ }\textbf {\bibinfo {volume} {130}},\
  \bibinfo {pages} {240801} (\bibinfo {year} {2023})}\BibitemShut {NoStop}%
\bibitem [{\citenamefont {Srivastava}\ \emph
  {et~al.}(2023{\natexlab{a}})\citenamefont {Srivastava}, \citenamefont
  {Horoshko},\ and\ \citenamefont {Kolobov}}]{Srivastava23b}%
  \BibitemOpen
  \bibfield  {author} {\bibinfo {author} {\bibfnamefont {S.}~\bibnamefont
  {Srivastava}}, \bibinfo {author} {\bibfnamefont {D.~B.}\ \bibnamefont
  {Horoshko}},\ and\ \bibinfo {author} {\bibfnamefont {M.~I.}\ \bibnamefont
  {Kolobov}},\ }\bibfield  {title} {\bibinfo {title} {Erecting time telescope
  for photonic quantum networks},\ }\href {https://doi.org/10.1364/OE.501609}
  {\bibfield  {journal} {\bibinfo  {journal} {Opt. Express}\ }\textbf {\bibinfo
  {volume} {31}},\ \bibinfo {pages} {38560} (\bibinfo {year}
  {2023}{\natexlab{a}})}\BibitemShut {NoStop}%
\bibitem [{\citenamefont {Patera}\ \emph {et~al.}(2023)\citenamefont {Patera},
  \citenamefont {Horoshko}, \citenamefont {Allgaier}, \citenamefont {Kolobov},\
  and\ \citenamefont {Silberhorn}}]{Patera23}%
  \BibitemOpen
  \bibfield  {author} {\bibinfo {author} {\bibfnamefont {G.}~\bibnamefont
  {Patera}}, \bibinfo {author} {\bibfnamefont {D.~B.}\ \bibnamefont
  {Horoshko}}, \bibinfo {author} {\bibfnamefont {M.}~\bibnamefont {Allgaier}},
  \bibinfo {author} {\bibfnamefont {M.~I.}\ \bibnamefont {Kolobov}},\ and\
  \bibinfo {author} {\bibfnamefont {C.}~\bibnamefont {Silberhorn}},\ }\bibfield
   {title} {\bibinfo {title} {Modal approach to quantum temporal imaging},\
  }\href {https://doi.org/10.1103/PhysRevA.108.043716} {\bibfield  {journal}
  {\bibinfo  {journal} {Phys. Rev. A}\ }\textbf {\bibinfo {volume} {108}},\
  \bibinfo {pages} {043716} (\bibinfo {year} {2023})}\BibitemShut {NoStop}%
\bibitem [{\citenamefont {Srivastava}\ \emph
  {et~al.}(2023{\natexlab{b}})\citenamefont {Srivastava}, \citenamefont
  {Horoshko},\ and\ \citenamefont {Kolobov}}]{Srivastava23}%
  \BibitemOpen
  \bibfield  {author} {\bibinfo {author} {\bibfnamefont {S.}~\bibnamefont
  {Srivastava}}, \bibinfo {author} {\bibfnamefont {D.~B.}\ \bibnamefont
  {Horoshko}},\ and\ \bibinfo {author} {\bibfnamefont {M.~I.}\ \bibnamefont
  {Kolobov}},\ }\bibfield  {title} {\bibinfo {title} {Making entangled photons
  indistinguishable by a time lens},\ }\href
  {https://doi.org/10.1103/PhysRevA.107.033705} {\bibfield  {journal} {\bibinfo
   {journal} {Phys. Rev. A}\ }\textbf {\bibinfo {volume} {107}},\ \bibinfo
  {pages} {033705} (\bibinfo {year} {2023}{\natexlab{b}})}\BibitemShut
  {NoStop}%
\bibitem [{\citenamefont {Quesada}\ and\ \citenamefont
  {Sipe}(2015)}]{Quesada15}%
  \BibitemOpen
  \bibfield  {author} {\bibinfo {author} {\bibfnamefont {N.}~\bibnamefont
  {Quesada}}\ and\ \bibinfo {author} {\bibfnamefont {J.~E.}\ \bibnamefont
  {Sipe}},\ }\bibfield  {title} {\bibinfo {title} {Time-ordering effects in the
  generation of entangled photons using nonlinear optical processes},\ }\href
  {https://doi.org/10.1103/PhysRevLett.114.093903} {\bibfield  {journal}
  {\bibinfo  {journal} {Phys. Rev. Lett.}\ }\textbf {\bibinfo {volume} {114}},\
  \bibinfo {pages} {093903} (\bibinfo {year} {2015})}\BibitemShut {NoStop}%
\bibitem [{\citenamefont {Christ}\ \emph {et~al.}(2013)\citenamefont {Christ},
  \citenamefont {Brecht}, \citenamefont {Mauerer},\ and\ \citenamefont
  {Silberhorn}}]{Christ13}%
  \BibitemOpen
  \bibfield  {author} {\bibinfo {author} {\bibfnamefont {A.}~\bibnamefont
  {Christ}}, \bibinfo {author} {\bibfnamefont {B.}~\bibnamefont {Brecht}},
  \bibinfo {author} {\bibfnamefont {W.}~\bibnamefont {Mauerer}},\ and\ \bibinfo
  {author} {\bibfnamefont {C.}~\bibnamefont {Silberhorn}},\ }\bibfield  {title}
  {\bibinfo {title} {Theory of quantum frequency conversion and type-{II}
  parametric down-conversion in the high-gain regime},\ }\href
  {http://stacks.iop.org/1367-2630/15/i=5/a=053038} {\bibfield  {journal}
  {\bibinfo  {journal} {New J. Phys.}\ }\textbf {\bibinfo {volume} {15}},\
  \bibinfo {pages} {053038} (\bibinfo {year} {2013})}\BibitemShut {NoStop}%
\bibitem [{\citenamefont {Lipfert}\ \emph {et~al.}(2018)\citenamefont
  {Lipfert}, \citenamefont {Horoshko}, \citenamefont {Patera},\ and\
  \citenamefont {Kolobov}}]{Lipfert18}%
  \BibitemOpen
  \bibfield  {author} {\bibinfo {author} {\bibfnamefont {T.}~\bibnamefont
  {Lipfert}}, \bibinfo {author} {\bibfnamefont {D.~B.}\ \bibnamefont
  {Horoshko}}, \bibinfo {author} {\bibfnamefont {G.}~\bibnamefont {Patera}},\
  and\ \bibinfo {author} {\bibfnamefont {M.~I.}\ \bibnamefont {Kolobov}},\
  }\bibfield  {title} {\bibinfo {title} {Bloch-{Messiah} decomposition and
  {Magnus} expansion for parametric down-conversion with monochromatic pump},\
  }\href {https://doi.org/10.1103/PhysRevA.98.013815} {\bibfield  {journal}
  {\bibinfo  {journal} {Phys. Rev. A}\ }\textbf {\bibinfo {volume} {98}},\
  \bibinfo {pages} {013815} (\bibinfo {year} {2018})}\BibitemShut {NoStop}%
\bibitem [{\citenamefont {Grice}\ \emph {et~al.}(2001)\citenamefont {Grice},
  \citenamefont {U'Ren},\ and\ \citenamefont {Walmsley}}]{Grice01}%
  \BibitemOpen
  \bibfield  {author} {\bibinfo {author} {\bibfnamefont {W.~P.}\ \bibnamefont
  {Grice}}, \bibinfo {author} {\bibfnamefont {A.~B.}\ \bibnamefont {U'Ren}},\
  and\ \bibinfo {author} {\bibfnamefont {I.~A.}\ \bibnamefont {Walmsley}},\
  }\bibfield  {title} {\bibinfo {title} {Eliminating frequency and space-time
  correlations in multiphoton states},\ }\href
  {https://doi.org/10.1103/PhysRevA.64.063815} {\bibfield  {journal} {\bibinfo
  {journal} {Phys. Rev. A}\ }\textbf {\bibinfo {volume} {64}},\ \bibinfo
  {pages} {063815} (\bibinfo {year} {2001})}\BibitemShut {NoStop}%
\bibitem [{\citenamefont {Horoshko}\ \emph
  {et~al.}(2019{\natexlab{a}})\citenamefont {Horoshko}, \citenamefont
  {La~Volpe}, \citenamefont {Arzani}, \citenamefont {Treps}, \citenamefont
  {Fabre},\ and\ \citenamefont {Kolobov}}]{Horoshko19}%
  \BibitemOpen
  \bibfield  {author} {\bibinfo {author} {\bibfnamefont {D.~B.}\ \bibnamefont
  {Horoshko}}, \bibinfo {author} {\bibfnamefont {L.}~\bibnamefont {La~Volpe}},
  \bibinfo {author} {\bibfnamefont {F.}~\bibnamefont {Arzani}}, \bibinfo
  {author} {\bibfnamefont {N.}~\bibnamefont {Treps}}, \bibinfo {author}
  {\bibfnamefont {C.}~\bibnamefont {Fabre}},\ and\ \bibinfo {author}
  {\bibfnamefont {M.~I.}\ \bibnamefont {Kolobov}},\ }\bibfield  {title}
  {\bibinfo {title} {Bloch-{Messiah} reduction for twin beams of light},\
  }\href {https://doi.org/10.1103/PhysRevA.100.013837} {\bibfield  {journal}
  {\bibinfo  {journal} {Phys. Rev. A}\ }\textbf {\bibinfo {volume} {100}},\
  \bibinfo {pages} {013837} (\bibinfo {year} {2019}{\natexlab{a}})}\BibitemShut
  {NoStop}%
\bibitem [{\citenamefont {Horoshko}\ and\ \citenamefont
  {Kolobov}(2024)}]{Horoshko23b}%
  \BibitemOpen
  \bibfield  {author} {\bibinfo {author} {\bibfnamefont {D.~B.}\ \bibnamefont
  {Horoshko}}\ and\ \bibinfo {author} {\bibfnamefont {M.~I.}\ \bibnamefont
  {Kolobov}},\ }\bibfield  {title} {\bibinfo {title} {Interferometric sorting
  of temporal {Hermite-Gauss} modes via temporal {Gouy} phase},\ }\href
  {https://doi.org/10.1103/PhysRevA.110.033721} {\bibfield  {journal} {\bibinfo
   {journal} {Phys. Rev. A}\ }\textbf {\bibinfo {volume} {110}},\ \bibinfo
  {pages} {033721} (\bibinfo {year} {2024})}\BibitemShut {NoStop}%
\bibitem [{\citenamefont {Law}\ and\ \citenamefont {Eberly}(2004)}]{Law04}%
  \BibitemOpen
  \bibfield  {author} {\bibinfo {author} {\bibfnamefont {C.~K.}\ \bibnamefont
  {Law}}\ and\ \bibinfo {author} {\bibfnamefont {J.~H.}\ \bibnamefont
  {Eberly}},\ }\bibfield  {title} {\bibinfo {title} {Analysis and
  interpretation of high transverse entanglement in optical parametric down
  conversion},\ }\href {https://doi.org/10.1103/PhysRevLett.92.127903}
  {\bibfield  {journal} {\bibinfo  {journal} {Phys. Rev. Lett.}\ }\textbf
  {\bibinfo {volume} {92}},\ \bibinfo {pages} {127903} (\bibinfo {year}
  {2004})}\BibitemShut {NoStop}%
\bibitem [{\citenamefont {Horoshko}\ \emph {et~al.}(2012)\citenamefont
  {Horoshko}, \citenamefont {Patera}, \citenamefont {Gatti},\ and\
  \citenamefont {Kolobov}}]{Horoshko12}%
  \BibitemOpen
  \bibfield  {author} {\bibinfo {author} {\bibfnamefont {D.~B.}\ \bibnamefont
  {Horoshko}}, \bibinfo {author} {\bibfnamefont {G.}~\bibnamefont {Patera}},
  \bibinfo {author} {\bibfnamefont {A.}~\bibnamefont {Gatti}},\ and\ \bibinfo
  {author} {\bibfnamefont {M.~I.}\ \bibnamefont {Kolobov}},\ }\bibfield
  {title} {\bibinfo {title} {X-entangled biphotons: Schmidt number for 2{D}
  model},\ }\href {https://doi.org/10.1140/epjd/e2012-30099-y} {\bibfield
  {journal} {\bibinfo  {journal} {Eur. Phys. J. D}\ }\textbf {\bibinfo {volume}
  {66}},\ \bibinfo {pages} {239} (\bibinfo {year} {2012})}\BibitemShut
  {NoStop}%
\bibitem [{\citenamefont {Gatti}\ \emph {et~al.}(2012)\citenamefont {Gatti},
  \citenamefont {Corti}, \citenamefont {Brambilla},\ and\ \citenamefont
  {Horoshko}}]{Gatti12}%
  \BibitemOpen
  \bibfield  {author} {\bibinfo {author} {\bibfnamefont {A.}~\bibnamefont
  {Gatti}}, \bibinfo {author} {\bibfnamefont {T.}~\bibnamefont {Corti}},
  \bibinfo {author} {\bibfnamefont {E.}~\bibnamefont {Brambilla}},\ and\
  \bibinfo {author} {\bibfnamefont {D.~B.}\ \bibnamefont {Horoshko}},\
  }\bibfield  {title} {\bibinfo {title} {Dimensionality of the spatiotemporal
  entanglement of parametric down-conversion photon pairs},\ }\href
  {https://doi.org/10.1103/PhysRevA.86.053803} {\bibfield  {journal} {\bibinfo
  {journal} {Phys. Rev. A}\ }\textbf {\bibinfo {volume} {86}},\ \bibinfo
  {pages} {053803} (\bibinfo {year} {2012})}\BibitemShut {NoStop}%
\bibitem [{\citenamefont {Sharapova}\ \emph {et~al.}(2020)\citenamefont
  {Sharapova}, \citenamefont {Frascella}, \citenamefont {Riabinin},
  \citenamefont {P\'erez}, \citenamefont {Tikhonova}, \citenamefont {Lemieux},
  \citenamefont {Boyd}, \citenamefont {Leuchs},\ and\ \citenamefont
  {Chekhova}}]{Sharapova20}%
  \BibitemOpen
  \bibfield  {author} {\bibinfo {author} {\bibfnamefont {P.~R.}\ \bibnamefont
  {Sharapova}}, \bibinfo {author} {\bibfnamefont {G.}~\bibnamefont
  {Frascella}}, \bibinfo {author} {\bibfnamefont {M.}~\bibnamefont {Riabinin}},
  \bibinfo {author} {\bibfnamefont {A.~M.}\ \bibnamefont {P\'erez}}, \bibinfo
  {author} {\bibfnamefont {O.~V.}\ \bibnamefont {Tikhonova}}, \bibinfo {author}
  {\bibfnamefont {S.}~\bibnamefont {Lemieux}}, \bibinfo {author} {\bibfnamefont
  {R.~W.}\ \bibnamefont {Boyd}}, \bibinfo {author} {\bibfnamefont
  {G.}~\bibnamefont {Leuchs}},\ and\ \bibinfo {author} {\bibfnamefont {M.~V.}\
  \bibnamefont {Chekhova}},\ }\bibfield  {title} {\bibinfo {title} {Properties
  of bright squeezed vacuum at increasing brightness},\ }\href
  {https://doi.org/10.1103/PhysRevResearch.2.013371} {\bibfield  {journal}
  {\bibinfo  {journal} {Phys. Rev. Res.}\ }\textbf {\bibinfo {volume} {2}},\
  \bibinfo {pages} {013371} (\bibinfo {year} {2020})}\BibitemShut {NoStop}%
\bibitem [{\citenamefont {Averchenko}\ \emph {et~al.}(2020)\citenamefont
  {Averchenko}, \citenamefont {Frascella}, \citenamefont {Kalash},
  \citenamefont {Cavanna},\ and\ \citenamefont {Chekhova}}]{Averchenko20}%
  \BibitemOpen
  \bibfield  {author} {\bibinfo {author} {\bibfnamefont {V.~A.}\ \bibnamefont
  {Averchenko}}, \bibinfo {author} {\bibfnamefont {G.}~\bibnamefont
  {Frascella}}, \bibinfo {author} {\bibfnamefont {M.}~\bibnamefont {Kalash}},
  \bibinfo {author} {\bibfnamefont {A.}~\bibnamefont {Cavanna}},\ and\ \bibinfo
  {author} {\bibfnamefont {M.~V.}\ \bibnamefont {Chekhova}},\ }\bibfield
  {title} {\bibinfo {title} {Reconstructing two-dimensional spatial modes for
  classical and quantum light},\ }\href
  {https://doi.org/10.1103/PhysRevA.102.053725} {\bibfield  {journal} {\bibinfo
   {journal} {Phys. Rev. A}\ }\textbf {\bibinfo {volume} {102}},\ \bibinfo
  {pages} {053725} (\bibinfo {year} {2020})}\BibitemShut {NoStop}%
\bibitem [{\citenamefont {Kopylov}\ \emph {et~al.}(2025)\citenamefont
  {Kopylov}, \citenamefont {Meier},\ and\ \citenamefont
  {Sharapova}}]{Kopylov25}%
  \BibitemOpen
  \bibfield  {author} {\bibinfo {author} {\bibfnamefont {D.~A.}\ \bibnamefont
  {Kopylov}}, \bibinfo {author} {\bibfnamefont {T.}~\bibnamefont {Meier}},\
  and\ \bibinfo {author} {\bibfnamefont {P.~R.}\ \bibnamefont {Sharapova}},\
  }\bibfield  {title} {\bibinfo {title} {Theory of multimode squeezed light
  generation in lossy media},\ }\href
  {https://doi.org/10.22331/q-2025-02-04-1621} {\bibfield  {journal} {\bibinfo
  {journal} {{Quantum}}\ }\textbf {\bibinfo {volume} {9}},\ \bibinfo {pages}
  {1621} (\bibinfo {year} {2025})}\BibitemShut {NoStop}%
\bibitem [{\citenamefont {Bobrov}\ \emph {et~al.}(2013)\citenamefont {Bobrov},
  \citenamefont {Straupe}, \citenamefont {Kovlakov},\ and\ \citenamefont
  {Kulik}}]{Bobrov13}%
  \BibitemOpen
  \bibfield  {author} {\bibinfo {author} {\bibfnamefont {I.~B.}\ \bibnamefont
  {Bobrov}}, \bibinfo {author} {\bibfnamefont {S.~S.}\ \bibnamefont {Straupe}},
  \bibinfo {author} {\bibfnamefont {E.~V.}\ \bibnamefont {Kovlakov}},\ and\
  \bibinfo {author} {\bibfnamefont {S.~P.}\ \bibnamefont {Kulik}},\ }\bibfield
  {title} {\bibinfo {title} {Schmidt-like coherent mode decomposition and
  spatial intensity correlations of thermal light},\ }\href
  {https://doi.org/10.1088/1367-2630/15/7/073016} {\bibfield  {journal}
  {\bibinfo  {journal} {New J. Phys.}\ }\textbf {\bibinfo {volume} {15}},\
  \bibinfo {pages} {073016} (\bibinfo {year} {2013})}\BibitemShut {NoStop}%
\bibitem [{\citenamefont {Ansari}\ \emph {et~al.}(2018)\citenamefont {Ansari},
  \citenamefont {Donohue}, \citenamefont {Brecht},\ and\ \citenamefont
  {Silberhorn}}]{Ansari18}%
  \BibitemOpen
  \bibfield  {author} {\bibinfo {author} {\bibfnamefont {V.}~\bibnamefont
  {Ansari}}, \bibinfo {author} {\bibfnamefont {J.~M.}\ \bibnamefont {Donohue}},
  \bibinfo {author} {\bibfnamefont {B.}~\bibnamefont {Brecht}},\ and\ \bibinfo
  {author} {\bibfnamefont {C.}~\bibnamefont {Silberhorn}},\ }\bibfield  {title}
  {\bibinfo {title} {Tailoring nonlinear processes for quantum optics with
  pulsed temporal-mode encodings},\ }\href
  {https://doi.org/10.1364/OPTICA.5.000534} {\bibfield  {journal} {\bibinfo
  {journal} {Optica}\ }\textbf {\bibinfo {volume} {5}},\ \bibinfo {pages} {534}
  (\bibinfo {year} {2018})}\BibitemShut {NoStop}%
\bibitem [{\citenamefont {Giovannetti}\ \emph {et~al.}(2002)\citenamefont
  {Giovannetti}, \citenamefont {Maccone}, \citenamefont {Shapiro},\ and\
  \citenamefont {Wong}}]{Giovannetti02}%
  \BibitemOpen
  \bibfield  {author} {\bibinfo {author} {\bibfnamefont {V.}~\bibnamefont
  {Giovannetti}}, \bibinfo {author} {\bibfnamefont {L.}~\bibnamefont
  {Maccone}}, \bibinfo {author} {\bibfnamefont {J.~H.}\ \bibnamefont
  {Shapiro}},\ and\ \bibinfo {author} {\bibfnamefont {F.~N.~C.}\ \bibnamefont
  {Wong}},\ }\bibfield  {title} {\bibinfo {title} {Generating entangled
  two-photon states with coincident frequencies},\ }\href
  {https://doi.org/10.1103/PhysRevLett.88.183602} {\bibfield  {journal}
  {\bibinfo  {journal} {Phys. Rev. Lett.}\ }\textbf {\bibinfo {volume} {88}},\
  \bibinfo {pages} {183602} (\bibinfo {year} {2002})}\BibitemShut {NoStop}%
\bibitem [{\citenamefont {Brecht}\ \emph {et~al.}(2015)\citenamefont {Brecht},
  \citenamefont {Reddy}, \citenamefont {Silberhorn},\ and\ \citenamefont
  {Raymer}}]{Brecht15}%
  \BibitemOpen
  \bibfield  {author} {\bibinfo {author} {\bibfnamefont {B.}~\bibnamefont
  {Brecht}}, \bibinfo {author} {\bibfnamefont {D.~V.}\ \bibnamefont {Reddy}},
  \bibinfo {author} {\bibfnamefont {C.}~\bibnamefont {Silberhorn}},\ and\
  \bibinfo {author} {\bibfnamefont {M.~G.}\ \bibnamefont {Raymer}},\ }\bibfield
   {title} {\bibinfo {title} {Photon temporal modes: A complete framework for
  quantum information science},\ }\href
  {https://doi.org/10.1103/PhysRevX.5.041017} {\bibfield  {journal} {\bibinfo
  {journal} {Phys. Rev. X}\ }\textbf {\bibinfo {volume} {5}},\ \bibinfo {pages}
  {041017} (\bibinfo {year} {2015})}\BibitemShut {NoStop}%
\bibitem [{\citenamefont {K\"onig}\ and\ \citenamefont {Wong}(2004)}]{Konig04}%
  \BibitemOpen
  \bibfield  {author} {\bibinfo {author} {\bibfnamefont {F.}~\bibnamefont
  {K\"onig}}\ and\ \bibinfo {author} {\bibfnamefont {F.~N.~C.}\ \bibnamefont
  {Wong}},\ }\bibfield  {title} {\bibinfo {title} {Extended phase matching of
  second-harmonic generation in periodically poled {KTiOPO4} with zero
  group-velocity mismatch},\ }\href {https://doi.org/10.1063/1.1668320}
  {\bibfield  {journal} {\bibinfo  {journal} {Appl. Phys. Lett.}\ }\textbf
  {\bibinfo {volume} {84}},\ \bibinfo {pages} {1644} (\bibinfo {year}
  {2004})}\BibitemShut {NoStop}%
\bibitem [{\citenamefont {Mosley}\ \emph {et~al.}(2008)\citenamefont {Mosley},
  \citenamefont {Lundeen}, \citenamefont {Smith}, \citenamefont {Wasylczyk},
  \citenamefont {U'Ren}, \citenamefont {Silberhorn},\ and\ \citenamefont
  {Walmsley}}]{Mosley08}%
  \BibitemOpen
  \bibfield  {author} {\bibinfo {author} {\bibfnamefont {P.~J.}\ \bibnamefont
  {Mosley}}, \bibinfo {author} {\bibfnamefont {J.~S.}\ \bibnamefont {Lundeen}},
  \bibinfo {author} {\bibfnamefont {B.~J.}\ \bibnamefont {Smith}}, \bibinfo
  {author} {\bibfnamefont {P.}~\bibnamefont {Wasylczyk}}, \bibinfo {author}
  {\bibfnamefont {A.~B.}\ \bibnamefont {U'Ren}}, \bibinfo {author}
  {\bibfnamefont {C.}~\bibnamefont {Silberhorn}},\ and\ \bibinfo {author}
  {\bibfnamefont {I.~A.}\ \bibnamefont {Walmsley}},\ }\bibfield  {title}
  {\bibinfo {title} {Heralded generation of ultrafast single photons in pure
  quantum states},\ }\href {https://doi.org/10.1103/PhysRevLett.100.133601}
  {\bibfield  {journal} {\bibinfo  {journal} {Phys. Rev. Lett.}\ }\textbf
  {\bibinfo {volume} {100}},\ \bibinfo {pages} {133601} (\bibinfo {year}
  {2008})}\BibitemShut {NoStop}%
\bibitem [{\citenamefont {Grangier}\ \emph {et~al.}(1986)\citenamefont
  {Grangier}, \citenamefont {Roger},\ and\ \citenamefont
  {Aspect}}]{Grangier86}%
  \BibitemOpen
  \bibfield  {author} {\bibinfo {author} {\bibfnamefont {P.}~\bibnamefont
  {Grangier}}, \bibinfo {author} {\bibfnamefont {G.}~\bibnamefont {Roger}},\
  and\ \bibinfo {author} {\bibfnamefont {A.}~\bibnamefont {Aspect}},\
  }\bibfield  {title} {\bibinfo {title} {Experimental evidence for a photon
  anticorrelation effect on a beam splitter: a new light on single-photon
  interferences},\ }\href@noop {} {\bibfield  {journal} {\bibinfo  {journal}
  {Europhys. Lett.}\ }\textbf {\bibinfo {volume} {1}},\ \bibinfo {pages} {173}
  (\bibinfo {year} {1986})}\BibitemShut {NoStop}%
\bibitem [{\citenamefont {U'Ren}\ \emph {et~al.}(2005)\citenamefont {U'Ren},
  \citenamefont {Silberhorn}, \citenamefont {Ball}, \citenamefont {Banaszek},\
  and\ \citenamefont {Walmsley}}]{URen05}%
  \BibitemOpen
  \bibfield  {author} {\bibinfo {author} {\bibfnamefont {A.~B.}\ \bibnamefont
  {U'Ren}}, \bibinfo {author} {\bibfnamefont {C.}~\bibnamefont {Silberhorn}},
  \bibinfo {author} {\bibfnamefont {J.~L.}\ \bibnamefont {Ball}}, \bibinfo
  {author} {\bibfnamefont {K.}~\bibnamefont {Banaszek}},\ and\ \bibinfo
  {author} {\bibfnamefont {I.~A.}\ \bibnamefont {Walmsley}},\ }\bibfield
  {title} {\bibinfo {title} {Characterization of the nonclassical nature of
  conditionally prepared single photons},\ }\href
  {https://doi.org/10.1103/PhysRevA.72.021802} {\bibfield  {journal} {\bibinfo
  {journal} {Phys. Rev. A}\ }\textbf {\bibinfo {volume} {72}},\ \bibinfo
  {pages} {021802} (\bibinfo {year} {2005})}\BibitemShut {NoStop}%
\bibitem [{\citenamefont {Bocquillon}\ \emph {et~al.}(2009)\citenamefont
  {Bocquillon}, \citenamefont {Couteau}, \citenamefont {Razavi}, \citenamefont
  {Laflamme},\ and\ \citenamefont {Weihs}}]{Bocquillon09}%
  \BibitemOpen
  \bibfield  {author} {\bibinfo {author} {\bibfnamefont {E.}~\bibnamefont
  {Bocquillon}}, \bibinfo {author} {\bibfnamefont {C.}~\bibnamefont {Couteau}},
  \bibinfo {author} {\bibfnamefont {M.}~\bibnamefont {Razavi}}, \bibinfo
  {author} {\bibfnamefont {R.}~\bibnamefont {Laflamme}},\ and\ \bibinfo
  {author} {\bibfnamefont {G.}~\bibnamefont {Weihs}},\ }\bibfield  {title}
  {\bibinfo {title} {Coherence measures for heralded single-photon sources},\
  }\href {https://doi.org/10.1103/PhysRevA.79.035801} {\bibfield  {journal}
  {\bibinfo  {journal} {Phys. Rev. A}\ }\textbf {\bibinfo {volume} {79}},\
  \bibinfo {pages} {035801} (\bibinfo {year} {2009})}\BibitemShut {NoStop}%
\bibitem [{\citenamefont {Bettelli}(2010)}]{Bettelli10}%
  \BibitemOpen
  \bibfield  {author} {\bibinfo {author} {\bibfnamefont {S.}~\bibnamefont
  {Bettelli}},\ }\bibfield  {title} {\bibinfo {title} {Comment on ``{Coherence}
  measures for heralded single-photon sources''},\ }\href
  {https://doi.org/10.1103/PhysRevA.81.037801} {\bibfield  {journal} {\bibinfo
  {journal} {Phys. Rev. A}\ }\textbf {\bibinfo {volume} {81}},\ \bibinfo
  {pages} {037801} (\bibinfo {year} {2010})}\BibitemShut {NoStop}%
\bibitem [{\citenamefont {Lee}(1995)}]{Lee95}%
  \BibitemOpen
  \bibfield  {author} {\bibinfo {author} {\bibfnamefont {C.~T.}\ \bibnamefont
  {Lee}},\ }\bibfield  {title} {\bibinfo {title} {Theorem on nonclassical
  states},\ }\href {https://doi.org/10.1103/PhysRevA.52.3374} {\bibfield
  {journal} {\bibinfo  {journal} {Phys. Rev. A}\ }\textbf {\bibinfo {volume}
  {52}},\ \bibinfo {pages} {3374} (\bibinfo {year} {1995})}\BibitemShut
  {NoStop}%
\bibitem [{\citenamefont {Horoshko}\ \emph
  {et~al.}(2019{\natexlab{b}})\citenamefont {Horoshko}, \citenamefont
  {De~Bi\`evre}, \citenamefont {Patera},\ and\ \citenamefont
  {Kolobov}}]{Horoshko19tds}%
  \BibitemOpen
  \bibfield  {author} {\bibinfo {author} {\bibfnamefont {D.~B.}\ \bibnamefont
  {Horoshko}}, \bibinfo {author} {\bibfnamefont {S.}~\bibnamefont
  {De~Bi\`evre}}, \bibinfo {author} {\bibfnamefont {G.}~\bibnamefont
  {Patera}},\ and\ \bibinfo {author} {\bibfnamefont {M.~I.}\ \bibnamefont
  {Kolobov}},\ }\bibfield  {title} {\bibinfo {title} {Thermal-difference states
  of light: Quantum states of heralded photons},\ }\href
  {https://doi.org/10.1103/PhysRevA.100.053831} {\bibfield  {journal} {\bibinfo
   {journal} {Phys. Rev. A}\ }\textbf {\bibinfo {volume} {100}},\ \bibinfo
  {pages} {053831} (\bibinfo {year} {2019}{\natexlab{b}})}\BibitemShut
  {NoStop}%
\end{thebibliography}%
\end{document}